\documentclass[11pt]{article}
\usepackage[textwidth=450pt,textheight=650pt]{geometry}
\usepackage[round,colon]{natbib}
\usepackage{amsmath}
\usepackage{amssymb}
\usepackage{bbold}
\usepackage{tabularx}
\usepackage{color}
\usepackage{graphicx}
\usepackage{xspace}
\usepackage{caption}
\usepackage{slashed}
\usepackage{cancel}
\usepackage{soul}
\usepackage{pdfsync}
\usepackage{marginnote}
\usepackage{yfonts}   
\usepackage{url}

\newcommand{\al}{\alpha}

\newcommand{\bpi}{{\bar{\pi}}}

\newcommand{{\mole}}{\mbox{\,mole}}
\newcommand{\Deg}{^{\circ}}

\definecolor{purple}{rgb}{0.6,0,0.6}

\DeclareGraphicsExtensions{.pdf}

\setcaptionwidth{.9\textwidth}
 
\begin{document}

\hspace{110mm} \makebox[3cm][t]{Article, Discoveries}
\vspace{5mm}

{\LARGE \bf Reciprocal Nucleopeptides as the Ancestral Darwinian}\\[5pt]
{\LARGE \bf Self-Replicator
}\\[5ex] 
{\large\bf Eleanor F. Banwell$^{1*}$,  Bernard Piette$^{2*}$, Anne Taormina$^{2}$ and Jonathan Heddle$^{1,3,\natural}$}\\[5ex]
\llap{$^1$~}RIKEN, Hirosawa 2-1, Wako-shi, Saitama, 351-0198 Japan\\[3ex]
\llap{$^2$~}Department for Mathematical Sciences, Durham University, South Road,
Durham DH1 3LE, United Kingdom\\[3ex]
\llap{$^3$~}Bionanoscience and Biochemistry Laboratory,
 Malopolska Centre of Biotechnology, Jagiellonian University 30-387, Krakow, Poland.\\[3ex]
 \llap{$^*$~}These authors contributed equally\\[1ex]
 \llap{$^{\natural}$~}{\em Corresponding author:} Heddle, J.G. (jonathan.heddle@uj.edu.pl)\\[3ex]

This is a draft version of the manuscript. The final version was published in 
{\it Molecular Biology and Evolution}, msx292 (2017) and is available in open 
access from \url{https://doi.org/10.1093/molbev/msx292}.

{\bf Abstract}\\[1ex]
Even the simplest organisms are too complex to have spontaneously arisen fully-formed, yet precursors to first life must have emerged {\em ab initio} from their environment. A watershed event was the appearance of the first entity capable of evolution: the Initial Darwinian Ancestor. Here we suggest that nucleopeptide reciprocal replicators could have carried out this important role and contend that this is the simplest way to explain extant replication systems in a mathematically consistent way.  We propose short nucleic-acid templates on which amino-acylated adapters assembled. Spatial localization drives peptide ligation from activated precursors to generate phosphodiester-bond-catalytic peptides. Comprising autocatalytic protein and nucleic acid sequences, this dynamical system links and unifies several previous hypotheses and  provides a plausible model for  the emergence of DNA and the operational code.\\


\vspace{3ex}

{\bf Introduction}\\[1ex]
In contrast to our good understanding of more recent evolution, we still lack a coherent and robust theory that adequately explains the initial appearance of life on Earth (abiogenesis). In order to be complete, an abiogenic theory must describe a path from simple molecules to the Last Universal Common Ancestor (LUCA), requiring only a gradual increase in complexity.

The watershed event in abiogenesis was the emergence of the Initial Darwinian Ancestor (IDA): the first self-replicator (ignoring dead ends) and ancestral to all life on Earth (\citealt{Yarus:2011}). Following the insights of von Neumann, who proposed the kinematic model of self-replication (\citealt{Kemeni:1955}), necessary features of such a replicator are: Storage of the information for how to build a replicator; A processor to interpret information and select parts; An instance of the replicator.

In order to be viable, any proposal for the IDA's structure must fit with spontaneous emergence from prebiotic geochemistry and principles of self-replication. Currently, the most dominant abiogenesis theory is the ``RNA world'', which posits that the IDA was a self-replicating ribozyme, i.e. an RNA-dependent RNA polymerase (\citealt{Cech:2012}). Although popular, this theory has problems (\citealt{Kurland:2010}). For example, while it is plausible that molecules with the necessary replication characteristics can exist, length requirements seem to make their spontaneous emergence from the primordial milieu unlikely, nor does the RNA world explain the appearance of the operational code (\citealt{Noller:2012, RobertsonJoyce:2012}). Furthermore, it invokes three exchanges of function between RNA and other molecules to explain the coupling of polynucleotide and protein biosynthesis, namely transfer of information storage capability to DNA and polymerase activity to protein as well as  gain of peptide synthesis ability. This seems
an implausible situation in which no extant molecule continues in the role it initially held. Others have posited peptide and nucleopeptide worlds as solutions, but to the best of our knowledge, no single theory has emerged that parsimoniously answers the biggest questions.

Here we build on several foregoing concepts to propose an alternative theory based around a nucleopeptide reciprocal replicator that uses its polynucleotide and peptide components according to their strengths, thus avoiding the need to explain later coupling. We advocate a view of the IDA as a dynamical system, i.e. a system of equations describing the changes that occur over time in the self-replicator  presented here,  and we demonstrate that such an entity is both mathematically consistent and complies with all the logical requirements for life. While necessarily wide in view we hope that this work will provide a useful framework for further investigation of this fundamental question.\\

\begin{figure}[!ht]
\begin{tabular}{lll}
\includegraphics[width=55mm, keepaspectratio]{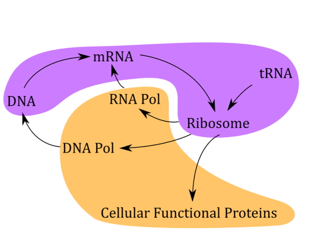}
&\includegraphics[width=55mm, keepaspectratio]{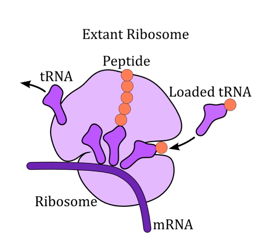}&
\includegraphics[width=45mm, keepaspectratio]{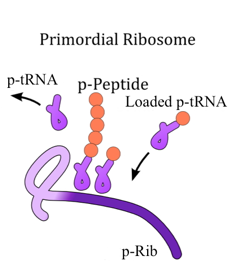}\\
\hskip 4cm(a)&\hskip 3cm(b)&\\[-8pt]
%
\\
\end{tabular}
\caption{\label{fig1ab}{\bf Replication Schemes}. (a) This simplified cellular replication schematic is common to all life today and likely reflects the ancestral form present in LUCA. Shading by molecule type (purple for nucleic acid and orange for protein), reveals a reciprocal nucleopeptide replicator. Although the ribosome is a large 
nucleoprotein complex, the catalytic centre has been shown to be a ribozyme (Moore and Steitz 2003) and so it is shaded purple in this scheme. (b) Comparison of the method of action of the extant ribosome with the proposed primordial analogue (components are shaded like for like). Today, tRNA molecules (mid purple) loaded with amino acids (orange) bind the mRNA (dark purple) in the ribosome (light purple), which co-ordinates and catalyses the peptidyl-transferase reaction. Although the present day modus operandi is regulated via far more complex interactions than the primordial version, the two schemes are fundamentally similar. Mixed nucleic acid structures, one performing a dual function as  primordial mRNA and  primordial ribosome (p-Rib) and a second functioning as a primordial tRNA (p-tRNA), provide a system wherein the former structure templates amino acid-loaded molecules of the latter.} 
\end{figure}

{\bf Model and Results}\\[1ex]
{\em Solving the chicken and egg problem}\\[1ex]
Given that any IDA must have been able to replicate in order to evolve, extant cellular replication machinery is an obvious source of clues to its identity. 
Common ancestry means that features shared by all life were part of LUCA.
By examining the common replication components present in LUCA, and then extrapolating further back to their simplest form, it is possible to reach a 
pre-LUCA, irreducibly complex, core replicator (Figure \ref{fig1ab}).

We see that in all cells, the required functions of a replicator are not carried out by a single molecule or even a single class of molecules, rather they are performed variously by nucleic acids (DNA, RNA) and proteins. When viewed by molecular class, the replicator has two components and is reciprocal in nature: polynucleotides rely on proteins for their polymerisation and vice versa. The question of which arose first is a chicken and egg conundrum that has dogged the field since the replication mechanisms were first elucidated (\citealt{GiriJain:2012}). In this work we suggest that, consistent with common ancestry and in contrast with the RNA world theory, the earliest replicator was a two - rather than a one - component system, composed of peptide and nucleic acids.\\[20pt]

{\em Assumptions of the model}\\[1ex]
We postulate that, in a nucleopeptide reciprocal replicator, the use of each component according to its strengths could deliver a viable IDA more compatible with evolution to LUCA replication machinery.  Although seemingly more complex than an individual replicating molecule, the resulting unified abiogenesis theory answers many hard questions and is ultimately more parsimonious.
In constructing our model, we make the following assumptions:
\begin{enumerate}
\item[{\em (i)}] {\em The existence of random sequences of short strands of mixed nucleic acids (XNA) likely consisting of ribonucleotides, deoxyribonucleotides and possibly other building blocks, as well as the existence  of random amino acids and short peptides produced abiotically.}

For this first assumption we have supposed a pool of interacting amino acids, nucleotides and related small molecules as well as a supply of metal ions, other inorganic catalysts and energy. A number of potential early earth conditions and reaction pathways resulting in these outcomes have been proposed, including the formamide reaction  (\citealt{Saladinoetal:2012a}) and cyanosulfidic chemistries (\citealt{Pateletal:2015}). Pools of pure molecules are unlikely; instead, mixtures would likely have comprised standard and non-standard amino acids as well as XNAs with mixed backbone architectures (\citealt{Trevinoetal:2011, Pinheiroetal:2012}). Such conditions would be conducive to the occasional spontaneous covalent attachment of nucleotides to each other to form longer polymer chains (\citealt{DaSilvaetal:2015}). 

\item[{\em (ii)}] {\em The existence of abiotically aminoacylated short XNA strands (primordial tRNAs (p-tRNAs))}

The second assumption is potentially troubling as amino acid activation is slow and thermodynamically unfavourable. However, amino acylation has been investigated in some detail and has been shown to be possible abiotically including, in some cases, the abiotic production of activated amino acids (\citealt{Illangasekareetal:1995, Lemanetal:2004, Giel-PietraszukBarciszewski:2006, Lehmannetal:2007,  Turketal:2010, Liuetal:2014}). Taken together these data suggest that multiple small amino-acylated tRNA-like primordial XNAs could have arisen. Though likely being XNA in nature, we refer to them as p-tRNA, reflecting their function. A similar nomenclature applies to p-Rib and p-mRNA.

\item[{\em (iii)}] {\em Conditions that allow a codon/anti-codon interaction between two or more charged p-tRNA for sufficient time and appropriate geometry to allow peptide bond formation, i.e. the functionality of a primordial ribosome (p-Rib)}

Our proposed p-Rib is an extreme simplification of the functionality of both the present day ribosome and mRNA (Figure 1). Initially, the p-Rib need only have been a  (close to) linear assembly template for the p-tRNAs to facilitate the peptidyl transferase reaction through an increase in local concentration. This mechanism is simple enough to emerge spontaneously and matches exactly the fundamental action of the extant ribosome  (Figure \ref{fig2}). The idea that a p-Rib may have an internal template rather than separate mRNA molecules and that an RNA 
strand could act as a way to bring charged tRNAs together has previously been suggested (\citealt{SchimmelHenderson:1994, WolfKoonin:2007, Morgens:2013}) and is known as an ``entropy trap'' (\citealt{Sievers:2004, Ruiz-Mirazo: 2014}). The concept has been demonstrated to be experimentally viable (\citealt{TamuraSchimmel:2003}) although in the latter case it is the primordial ribosomal rRNA strand itself that provides one of the two reacting amino acids. 

\vskip 0.5cm
 \begin{minipage}{\linewidth}
            \centering
            \includegraphics[width=6.5cm]{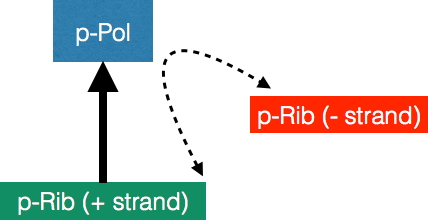}
            \captionof{figure}{\label{fig2}{\bf Models of primitive polymerization reactions}. 
             An XNA strand can function like a primordial ribosome (p-Rib) whereby one strand (+ strand) can template the production of a primordial polymerase (p-Pol) as indicated by the solid arrow. The action of this p-Pol is represented by the double-headed dotted arrow whereby it acts on the p-Rib (+ strand) to catalyse synthesis of the complementary sequence (- strand) and also on the - strand to produce more of the + strand.}
        \end{minipage}
\vskip 0.5cm
A functional operational system requires preferential charging of particular p-tRNAs to specific amino acids. Although there is evidence for such relationships in the stereochemical theory (\citealt{Woese:1965, Yarusetal:2009}), so far unequivocal proof has been elusive (\citealt{Yarusetal:2005b, KooninNovozhilov: 2009}). However, there is sufficient evidence to suggest at least a separation along grounds of hydrophobicity and charge using just a two-base codon (\citealt{KnightLandweber:2000, Biroetal:2003,  Rodinetal:2011}). Furthermore only a reduced set of amino acids (\citealt{Angyanetal:2014}) - possibly as few as four (\citealt{Ikehara:2002}) - need to have been provided in this way. The ``statistical protein'' hypothesis proposes that such a weak separation may have been sufficient to produce populations of active peptides (\citealt{Ikehara:2005, Vetsigianetal:2006}). Such ``primordial polymerases'' (p-Pol) need only have been small (see below) and spontaneous emergence of a template coding loosely for such a sequence seems plausible. The failure rate of such syntheses would be high but a p-Rib using the outlined primordial operational code to produce statistical p-Pol peptides could have been accurate enough to ensure its own survival.

\item[{\em (iv)}] {\em The viability of a very short peptide sequence to function as an RNA-dependent RNA polymerase}

Templated ligation is often proposed as a primordial self-replication mechanism, particularly for primitive replication of nucleic acid in RNA world type scenarios. However, these are associated with a number of problems as mentioned earlier. In addition, extant RNA/DNA synthesis proceeds via terminal elongation  (\citealt{PaulJoyce:2004, VidonnePhilp:2009}). To be consistent with the mechanism present in LUCA and pre-LUCA, the p-Pol should, preferably, have used a similar process.

During templated ligation, a parent molecule binds and ligates short substrates that must then dissociate to allow further access, but the product has greater binding affinity than the substrates and dissociation is slow. This product inhibition results in parabolic growth and limits the usefulness of templated ligation for replication (\citealt{IssacChmielewski:2002}). Conversely, in 1D sliding (or more accurately jumping), the catalyst may dock anywhere along a linear substrate and then diffuse by ``hops'' randomly in either direction until it reaches the reaction site; a successful ligation reaction has little impact on binding affinity and leaves the catalyst proximal to the next site. For simplicity our model assumes a single binding event between p-Pol and p-Rib followed by multiple polymerization events. A p-Pol proceeding via 1D sliding could catalyze phosphodiester bond formation between nucleotides bound by Watson and Crick base-pairing to a complementary XNA strand. Because p-Pol activity would be independent of substrate length, a relatively small catalyst could have acted on XNAs of considerable size. From inspection of present day polymerases such a peptide may have included sequences such as DxDGD  and/or GDD known to be conserved in their active sites and consisting of the amino acids thought to be amongst the very earliest in life (\citealt{Iyeretal:2003} \citealt{Koonin:1991}). 

In our simple system any such p-Pol must be very short to have any realistic chance of being produced by the primitive components described.  We must therefore ask if there is evidence  that small ({\it e.g.} less than 11 amino acid) peptides can have such a catalytic activity. Catalytic activity in general has been demonstrated for molecules as small as dipeptides (\citealt{Kochavietal:1997}). For polymerase activity in particular, it is known that randomly produced tripeptides can bind tightly and specifically to nucleotides (\citealt{Schneideretal:2000, McCleskeyetal:2003}). We suggest that a small peptide could arise with the ability to bind divalent metal ions, p-Rib and incoming nucleotides. It is interesting to note that small peptides can assemble into large and complex structures (\citealt{Bromleyetal:2008, Fletcheretal:2013}) with potentially sophisticated functionality: di-and tripeptides can self-assemble into larger nanotubes and intriguingly it has even been suggested that these structures could have acted as primitive RNA polymerases (\citealt{CarnyGazit:2005}).

In summary, the essence of the model is that on geological timescales, short linear polynucleotides may have been sufficient to template similar base-pairing interactions to those seen in the modern ribosome with small amino-acylated adapters. Given that the majority of ribosome activity stems from accurate substrate positioning, such templating could be sufficient to catalyze peptide bond formation and to deliver phosphodiester-bond-catalytic peptides. As backbone ligation reactions are unrelated to polynucleotide sequence, these generated primordial enzymes could have acted on a large subset of the available nucleic acid substrates, in turn producing more polynucleotide templates and resulting in an autocatalytic system.
\end{enumerate}
	
\vskip 1cm
{\em Mathematical Model}	

The IDA described above is attractive both for its simplicity and continuity with the existing mixed (protein/nucleic acid) replicator system in extant cells. However, the question remains as to whether such a system is mathematically consistent, could avoid collapse and instead become self-sustaining.  The number of parameters and variables needed to analyse the system in its full complexity is such that one is led to consider simplified models which nevertheless capture essential  features of interest. Here we consider a simple model of RNA-protein self-replication.

\begin{enumerate}
\item[{\em (a)}] {\em Constituents}

The main constituents of the simplest model of XNA-protein self-replication considered here  (see also Figure 1b and Figure 2) are a pool of free nucleotides and amino acids, polypeptide chains - including a family of polymerases -  and  polynucleotide chains as well as primordial tRNAs (p-tRNA) loaded with  single amino acids.

We introduce some notations. Generically, we consider polymer chains $\Pi$ made of $n$
{\em types} of building blocks labelled $1, \ldots , n$. In our models, the polymer chains are polypeptides and polynucleotides, and the building blocks are amino acids and codons respectively.  With a slight abuse of language, we call the number of constituents (building blocks) of a polymer chain its {\em length}. So hereafter, `lengths' are dimensionless.  The order in which these constituents appear in any chain is biologically significant, and we encode this information in finite ordered sequences of arbitrary length $L$ denoted $S\{L\} =(s_1, s_2,\ldots , s_L)$, whose elements $s_j,\,j=1,\ldots L$ label  the building blocks  forming the chains, in the order indicated in the sequences. Each element $s_j$ in the sequence $S\{L\}$ is an integer in the set $\{1,\ldots,n\}$ which refers to the type of building block occupying position $j$ in the chain.
There are therefore $n^L$ sequences of length $L$ if the model allows $n$ types of building blocks.
For instance, the sequence $S\{5\}=(1, 4, 3, 1, 3)$  in a model with, say, $n=4$ types of building blocks (amino acids or codons), corresponds to a polymer chain of length 5 whose first component is a type 1 building block, the second component is a type 4 and so on.
Given a sequence $S\{L\}$, we introduce subsequences $S\{L,\, j\}=(s_1, s_2,\ldots s_j)$ (resp.
$\widehat{S\{L,\, j\}}=(s_{L-j+1}, s_{L-j+2},\ldots s_{L})$), $ j=1,\ldots L$, whose elements are the  $j$ leftmost (resp. rightmost) elements of $S\{L\}$.
In particular, $S\{L,\,L\}\equiv \widehat{S\{L,\,L\}}\equiv S\{L\}$, $S\{L,\,1\}=s_1$ and $\widehat{S\{L,1\}}=s_L$.
We write 
$$
S\{L\} = (S\{L,\, L-\ell\},\widehat{S\{L,\,\ell\}}),\qquad 0 < \ell < L.
$$
In what follows we  sometimes refer to  families of polymer chains differing only by their length and obtained by 
removing some rightmost building blocks  from a chain of maximum 
length $L_{\rm {max}}$. Denoting by $\Pi_\ell^{S}$ a polymer chain of length $\ell$ and sequence $S\{\ell\}$ or
subsequence $S\{L,\ell\}$, both having $\ell$ elements with $L>\ell$, 
the family of polymer chains obtained from a chain of maximal length $L_{\rm {max}}$ and sequence $S\{L_{\rm max}\}$
 is given by $\left\{\Pi_{\ell}^{S}\right\}_{\ell =1, 2,\ldots L_{\rm {max}}}$.

In the specific case of  XNA/polynucleotide chains entering our model, we use $\Pi=R$  and the sequences are generically labelled as $\alpha\{\ell\}$. Their elements correspond to types of codons, and the complementary codon sequences  in the sense of  nucleic acids complementarity are $\overline{\alpha}{\{\ell\}}$.
Therefore a large class of XNA strands of length $\ell$ and sequence $\alpha\{\ell\}$ are denoted by $R_\ell^{\alpha}$, and in particular, $R_1^{\alpha_1}$ is a codon of type $\alpha_1$. Besides the generic sequences $\alpha\{\ell\}$ introduced above, a sequence denoted $\pi\{L_{\rm max}\}$, together with its subsequences $\pi\{L_{\rm max},\,\ell\}$ and $\widehat{\pi\{L_{\rm max},\,\ell\}}$ for $\ell =1,\ldots L_{\rm max}$ play a specific role: they correspond to polynucleotide chains that template the polymerisation of  a family of primordial peptide polymerases (p-Pol) through a process described in the next subsection, see also Figure \ref{fig_Prot_Polym}.
Using $\Pi=P$ to denote polypeptide chains, this family of polymerases derived 
from  $P_{L_{\rm max}}$ of maximal 
length $L_{\rm max}$, is $\left\{P_{\ell}^{\pi}\right\}_{\ell=2,\ldots,L_{\rm max}}$. These polymerases are such that 
$P_{\ell}^\pi=P_{\ell-1}^{\pi}+P_1^{\pi_{\ell}}$, with $P_1^{\pi_{\ell}}$ an amino acid $\pi_\ell$. We use the notation $P^{\pi}$ for a generic polymerase in the family. Alongside these polymerases, generic polypeptide
chains of length $\ell$ and sequence $\alpha\{\ell\}$ are labelled as $P_\ell^\alpha$. Proteins of length 1, $P_1^{\alpha_1}$, are single amino acids of type $\al_1$.


\item[{\em (b)}] {\em RNA-Protein replication scenario}

The scenario relies on three types of mechanisms:
\begin{itemize}
\item[(A)] The {\em spontaneous} polymerisation of polynucleotide and polypeptide chains, assumed to occur at a very slow rate, and their depolymerisation through being cleaved in two anywhere along the chains at a rate independent of where the cut occurs.
\item[(B)] The {\em non-spontaneous} polypeptide polymerisation occurring through a polynucleotide chain $R_L^S$ on which several p-tRNA molecules loaded with an amino acid dock and 
progressively build the polypeptide chain. More precisely, each codon of type $s$ of the polynucleotide 
chain binds with a p-tRNA, itself linked to an amino acid of type $s$. Note that we assume the same number $n$ of types of codons and amino acids.  This leads to a 
chain of amino acids matching the codon sequence $S\{L\}$ of the polynucleotide chain. 
The process is illustrated in Figure \ref{fig_Prot_Polym} for a polypeptide chain of length $L=4$ and amino acid sequence $S\{4\}=(s_1,s_2,s_3,s_4)$.
\item[(C)] The duplication of a polynucleotide chain $R_L^S$, 
of length $L \ge \ell_{\pi {\rm min}}$, as a two-step process. In the 
first step, a polypeptide polymerase  $P^\pi$, obtained by polymerisation via mechanism (B) using a polynucleotide $R_L^\pi$, scans the polynucleotide 
chain $R_L^S$ to generate its complementary polynucleotide chain $R_L^{\overline{S}}$. This is shown in Figure \ref{fig_RNA_polym}. The resulting polynucleotide chain $R_L^{\overline{S}}$ is then used to generate a copy of the original polynucleotide chain $R_L^S$ via the same mechanism (C).
\end{itemize}
\begin{figure}[ht!]
\begin{center}
\includegraphics[width=7.5cm, keepaspectratio]{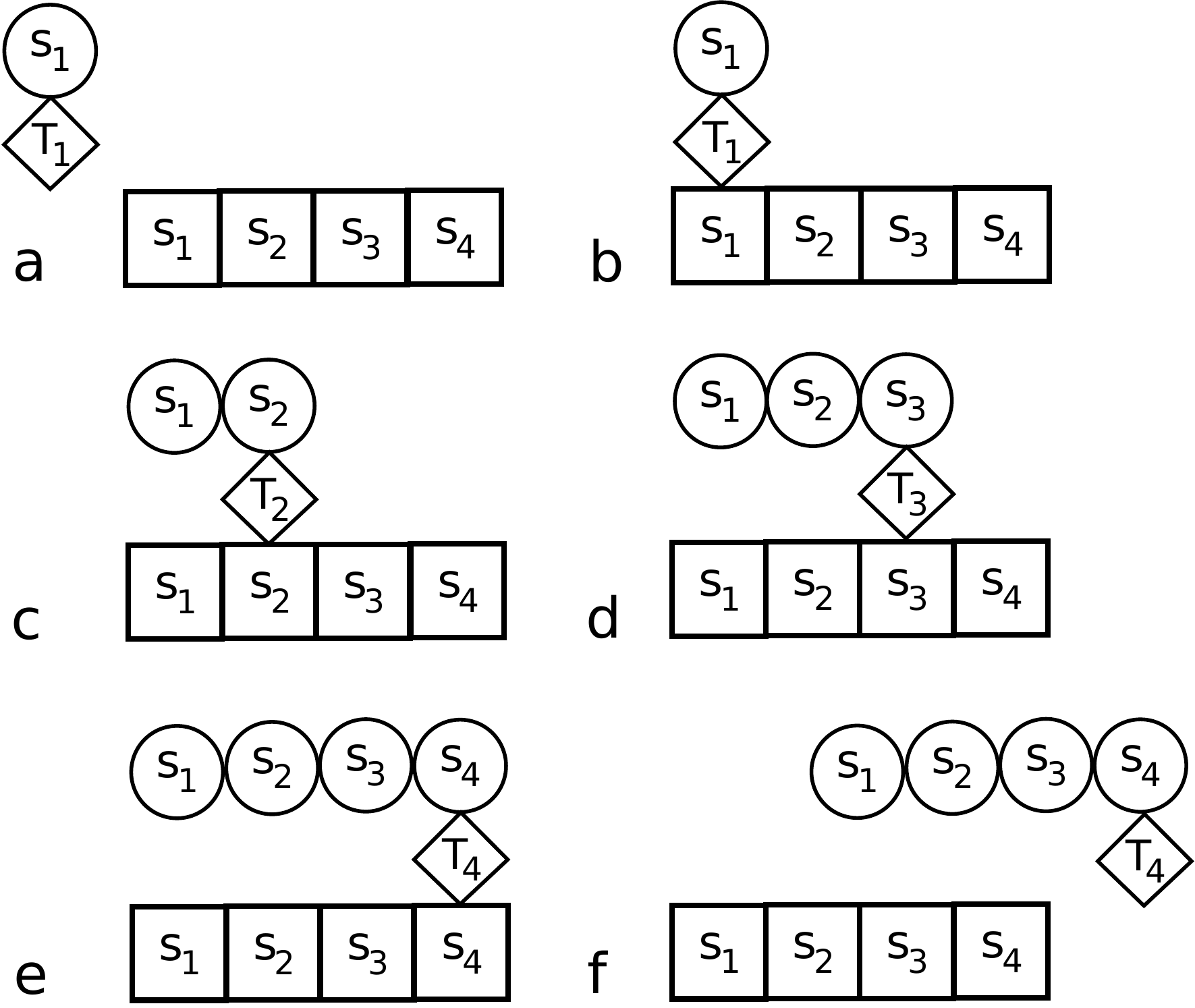}
\caption{\label{fig_Prot_Polym}
Mechanism (B): Polypeptide polymerisation in our model. The square boxes represent the codons of a polynucleotide chain (here, of length $L=4$) and the circles represent amino acids. The p-tRNA molecules are labelled $T_1, ..., T_4$.}
\end{center}
\end{figure}
\begin{figure}[ht!]
\begin{center}
\includegraphics[width=6.5cm, keepaspectratio]{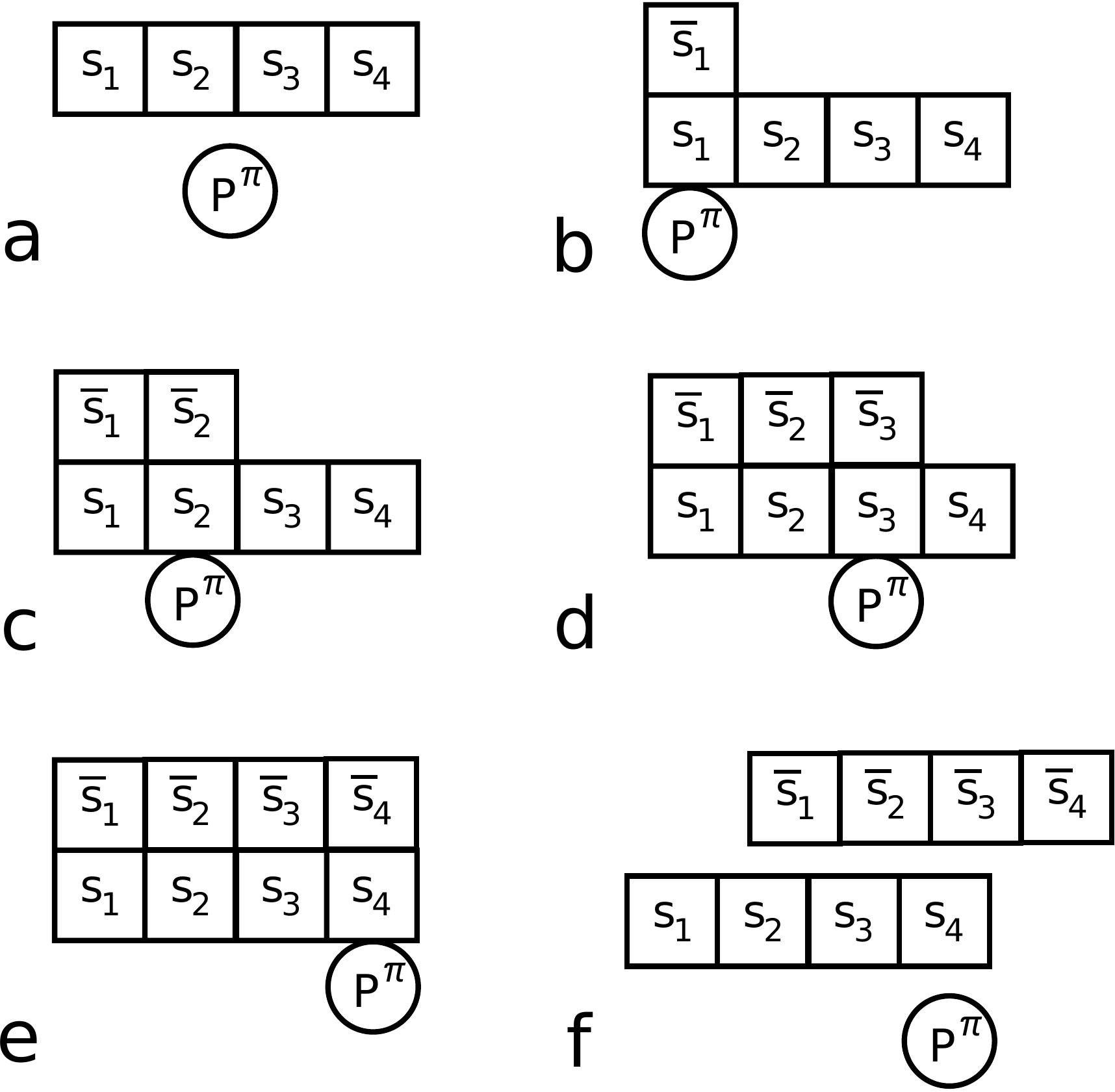}
\caption{\label{fig_RNA_polym}
First phase of Mechanism (C): Polymerisation of the complementary polynucleotide chain $R_L^{\overline{S}}$ 
catalysed by a primordial polymerase $P^{\pi}$.}
\end{center}
\end{figure}
The replicator crudely operates as follows:
\begin{itemize}
\item[-] Mechanism (A) provides a small pool of  polymer chains; among them, one finds short strands of XNA
with dual function (p-mRNA and p-Rib)
\item[-] Mechanism (B) provides 
polypeptide chains, including the polymerases
 (p-Pol, called $P^\pi$ here), by using the XNA produced through Mechanism (A) and 
Mechanism (C)
\item[-] $P^\pi$ are involved, through Mechanism (C), in the duplication of polynucleotides present in the environment, including the strands of XNA that participate in the very production of $P^\pi$ 
\end{itemize}

\item[{\em (c)}] {\em Reactions driving the replication and physical parameters}

For simplicity, we consider the 
polymerisation of polypeptide chains and the duplication of polynucleotide chains 
as single reactions where the reaction 
rates take into account all sub-processes as well as failure rates.

This leads to the following schematic reactions:
\begin{eqnarray}
{\rm \underline{Mechanism\,\, (A)}}\hskip 3cm&&\nonumber\\
R_L^{S} + R_1^{s_{L+1}} &\longrightarrow& R_{L+1}^{S}
\label{eq_polR}
\\
R_L^{S} &\longrightarrow& R_{L_-\ell}^{S} + R_{\ell}^{\widehat{S}}
\qquad \ell=1,\dots,L-1
\label{eq_depolR}
\\
P_L^{S} + P_1^{s_{L+1}} &\longrightarrow& P_{L+1}^{S} 
\label{eq_polP}
\\
P_L^{S} &\longrightarrow& P_{L_-\ell}^{S} + P_{\ell}^{\widehat{S}}
\qquad \ell=1,\dots,L-1
\label{eq_depolP}
\\
{\rm \underline{Mechanism\,\, (B)}}\hskip 3cm&&\nonumber\\
R_L^S + L\times TRP &\longrightarrow&  R_L^S + P_L^{S} 
\label{eq_dupPa}\\
{\rm \underline{Mechanism\,\, (C)}}\hskip 3cm&&\nonumber\\
R_L^{S} + L\times R_1 &\stackrel{P^{\pi}}{\longrightarrow}& 
        R_L^S + R_L^{\overline{S}} 
\label{eq_dupR}
\end{eqnarray}
where $TRP$ denotes  primordial tRNA loaded with a single 
amino acid.

The parameters for these reactions are (see the Supplementary Information 
for more details on the estimation of the parameter values):
\begin{itemize}
\item $K_R^+$ : polymerisation rate of polynucleotide chains 
[Equation \eqref{eq_polR}]; 
we have estimated the catalysed XNA polymerisation rate to be 
$4.2 \times 10^{-7}\,{\rm mol}^{-1}\,{\rm m}^3\,{\rm s}^{-1}$.
\item $K_R^-$ : depolymerisation rate of polynucleotide chains (hydrolysis)
[Equation \eqref{eq_depolR}]; taken to be
$8 \times 10^{-9}{\rm s}^{-1}$.
\item $K_P^+$ : polymerisation rate of polypeptide chains
[Equation \eqref{eq_polP}];  we have estimated it to be $2.8 \times 10^{-21}\,{\rm mol}^{-1}\,{\rm m}^3\,{\rm s}^{-1}$.
\item $K_{P,S,L}^-$ : depolymerisation rate of polypeptide chains of length $L$ and sequence $S$
[Equation \eqref{eq_depolP}];  we have estimated it to be in the range $4 \times 10^{-11}\,{\rm s}^{-1}- 5.1 \times 10^{-6}\,{\rm s}^{-1}$.
\item $k_{P,\,L}^+$ : polymerisation rate of a  polypeptide of length $L$ 
from the corresponding polynucleotide chain [Equation \eqref{eq_dupPa}]. It is reasonable to assume that 
$k_{P,L}^+=k_{P,1}^+/L$ 
and we have estimated $k_{P,\,1}^+$ to be  
$0.1\,{\rm mol}^{-1}\,{\rm m}^3\,{\rm s}^{-1}$.
\item $Z$ : the rate at which a polymerase  attaches 
  to a polynucleotide chain  [Equation \eqref{eq_dupR}] which we have estimated to 
be  $10^6\,{\rm mol}^{-1}\,{\rm m}^3\,{\rm s}^{-1}$.
\item $h_{R}$: the rate of attachment of a free polynucleotide to a 
  polynucleotide chain attached to a  p-Pol [Equation \eqref{eq_dupR}]. 
  We have estimated it to be  $10^{6}\,{\rm mol}^{-1}\,{\rm m}^3\,{\rm s}^{-1}$.
\item $k_{{\rm step}}$: the rate at which a polymerase moves by one step on the
polynucleotide [Equation \eqref{eq_dupR}]. We have estimated it to be in the range $2\times 10^{-2}\,{\rm s}^{-1}$
to  $4\times 10^{-5}\,{\rm s}^{-1}$.
 \end{itemize}
We now argue that the three parameters $Z,  h_{R}$ and $k_{{\rm step}}$ enter the dynamical system for the polymer concentrations in our model as two {\em physical} combinations denoted ${\cal K}(L)$ and ${\cal P}_b$ that we describe below.

First recall that we assume the existence of a pool of
nucleotides, amino acids and p-tRNA. 
The amount of {\em free} nucleotides and amino acids
is taken to be the difference between the total amount of these 
molecules and the total
amount of the corresponding polymerised material, ensuring total conservation.

We denote the concentration of polypeptide and polynucleotide 
chains respectively by $P_L^\al, P_L^\pi, P_L^\bpi $ and 
$R_L^{\al},R_L^{\pi}, R_L^{\bpi}$, all expressed in $\rm{mol}\,\rm{m}^{-3}$.
In particular, $P_1$ and $R_1$ are the concentrations of each type of free 
amino acids and
nucleotides respectively, and we  assume, for simplicity, that all types of 
amino acids/codons are equally available.

We also assume that the amount of loaded p-tRNA, $C_{\operatorname{p-tRNA}}$, remains 
proportional to the amount of free amino acids and that the concentration of 
p-tRNA is larger than $P_1$ so that most amino acids are loaded on a p-tRNA.
With these conventions, one has
\begin{equation}\label{ptRNAP1}
C_{\operatorname{p-tRNA}} = k_t P_1\qquad {\rm with}\qquad  k_t\approx 1.
\end{equation}

 {\bf {\em{Total reaction rate ${\cal K}(L)$ of polynucleotide polymerisation}}}
 
 If a complex reaction is the result of one event at rate $K$, and $m$ other,
identical, events at rate $k$, the average time to complete the reaction 
is the sum of the average times for each event. Hence the 
reaction rate is given by
\begin{eqnarray*}
\widetilde{{\cal K}}(K,k,m) &=& \left(\frac{1}{K}+\frac{m}{k}\right)^{-1}=\frac{Kk}{mK+k}.
\label{eq_Gsat}
\end{eqnarray*}

One such complex reaction in our model is the polymerisation of a
polynucleotide chain of length $L$, say, from its complementary chain (second phase of Mechanism (C)). Polymerases are characterised by the polymerising efficiency which, we assume, increases with $\ell$, up to $L_{\rm max}$. The first step in polymerisation requires a polymerase to  
attach itself 
to the template polynucleotide. This is only possible if the template polynucleotide has a minimum length, which we assume to be $\ell_{\pi{\rm min}}$. In the following, we assume that polymerases can polymerise polynucleotide chains of any length greater or equal to $\ell_{\pi{\rm min}}$. The corresponding reaction rate is given by
$Z\,P_\ell^{\pi}$ for a polymerase of length $\ell \ge \ell_{\pi{\rm min}}$.

The free nucleotides must then attach themselves to the 
polynucleotide-polymerase complex and the polymerase must move one step along 
the polynucleotide. The rate for each of these $L$ steps is
\begin{equation*} 
k_{R+} = \frac{k_{{\rm step}}\, h_{R} R_1}{k_{{\rm step}}+h_{R} R_1},
\end{equation*} 
and hence,
the rate of  polymerisation for a polynucleotide of length $L$ and polymerase of length $\ell$ is 
$\widetilde{{\cal K}}(Z\,P_\ell^{\pi},\,k_{R+},\,L)$. However,  it is assumed that 
 polymerases of several lengths are available and therefore, 
the total rate is given by
\begin{eqnarray*}
{\cal K}(L) &=& 
\left \{ \begin{array}{ll}
\sum_{\ell=\ell_{\pi{\rm min}}}^{L_{{\rm max}}} \widetilde{{\cal K}}(ZP_\ell^{\pi},\,k_{R+},\,L)\, W_\ell, & L \ge \ell_{\pi{\rm min}} \\
0 &  L < \ell_{\pi{\rm min}},\\
\end{array}
\right.
\label{eq_Ppolym}
\end{eqnarray*}
where  it is understood that $\ell_{\pi{\rm  min}}$ is the lower bound length 
for polymerase activity and $W_\ell$ is a quality factor given by
\begin{eqnarray*}
W_{\ell} = 
\left \{ \begin{array}{ll}
\displaystyle{\frac{\ell-\ell_{\pi{\rm min}}+1}{\ell_{\pi{\rm  max}}-\ell_{\pi{\rm  min}}+1}}\,  & \ell_{\pi{\rm min}} \le \ell \le \ell_{\pi{\rm  max}}\\
1 &  \ell_{\pi{\rm max}}  < \ell \le L_{{\rm max}}.\\
\end{array}
\right.
\label{eq_kRnl}
\end{eqnarray*}
Indeed, 
we expect long polymerases to be more efficient, so $W_{\ell}$ is taken 
to increase
with $\ell$ in the range $\ell_{\pi{\rm  min}} \le \ell \le \ell_{\pi{\rm  max}}$,
while polymerases of length $\ell > \ell_{\pi{\rm  max}}$ have the same level of 
activity as those with length $\ell = \ell_{\pi{\rm  max}}$ 
i.e. $W_{\ell \,>\, \ell_{{\rm max}}}=1$.

To avoid proliferation of parameters in our simulations, we have taken $\ell_{\pi{\rm max}}  = L_{{\rm max}}$, where $L_{\rm max}$ is the maximal polynucleotide chain's length.\\

{\bf {\em{Binding probability $\bold{{\cal P}_b}$ of a polynucleotide and a polymerase of length $L$  }}}
 
First note that it takes $L$ times longer to synthesise a polypeptide chain of length $L$  from its corresponding polynucleotide chain  than it  takes for one amino acid  to
bind itself to the polynucleotide.  The rate  is thus given by
$k_{P,\,L}^+\,P_1=(k_{P,\,1}^+/L)\,P_1$.

We now offer some considerations on depolymerisation. We assume that if a polymer $\Pi_L^S$  depolymerises, it does so by (potentially consecutive) cleavings. In the first step,  $\Pi_L^S$ can cleave in $L-1$ different positions, resulting in two smaller chains $L_1, L_2$ with $L=L_1+L_2$ and $1 \le L_{1,2} \le L-1$. This is the origin of the  factor
$(L-1)$ in the terms describing the depolymerisation of polymer chains in the dynamical systems equations presented in the next subsection. 

The concentration variations resulting from such depolymerisations must be 
carefully evaluated. 
A polymer $\Pi_L^S$ of length $L$ and sequence $S$, where $S$ 
 stands for any of $\alpha$, $\pi$ or $\bpi$, 
can be obtained by cleaving a polymer $\Pi_\ell^{\widetilde{S}}$  of length $\ell\,>\,L$ and sequence $\widetilde{S}=(S, T)$ 
where $T$ is a sequence of length $\ell-L$. Similarly it can be obtained by cleaving $\Pi_\ell^{\widetilde{S}^\prime}$ of sequence $\widetilde{S}^\prime=(T^\prime, S)$ where $T^\prime$ is also of length $\ell-L$.
If the rate of cleaving, $K^-_\Pi$, is assumed to be independent of the 
polymer length,  and since there are
$n^{\ell-L}$ different sequences $T$ and $T^\prime$, where $n$ is the number of amino acid or codon {\em types}, the rate of concentration variation of polymers of 
length $L$ resulting from the  depolymerisation of longer polymers is
\begin{equation}\label{rateconcvar}
\sum_{\ell=L+1}^{L_{{\rm max}}} n^{\ell-L}K_{\Pi}^-\,\Pi_{\ell}^{\widetilde{S}}\,+\, \sum_{\ell=L+1}^{L_{{\rm max}}} n^{\ell-L}K_{\Pi}^-\,\Pi_{\ell}^{\widetilde{S}^\prime}.
\end{equation}
Recall that we use the same notation for the concentration of a polymer of sequence $S$ and length $L$ and the polymer itself, namely  $\Pi^S_L$, and $\Pi$ is supposed to be set to $\Pi=P$ or $\Pi=R$ in our model. As already stressed, we assume polymers have at most length $L_{\rm max}$. Finally, when the concentrations $\Pi_L^{\widetilde{S}}$ and $\Pi_L^{\widetilde{S}^\prime}$ are equal, [Equation \eqref{rateconcvar}] can be rewritten as
\begin{equation*}
2\sum_{\ell=L+1}^{L_{{\rm max}}} n^{\ell-L}K_{\Pi}^-\,\Pi_{\ell}^{\widetilde{S}}.
\end{equation*}



The depolymerisation of polymerase $P_L^\pi$ requires special treatment. When  $P_L^\pi$ depolymerises, it generates a polymerase $P_{\ell}^\pi$ with $\ell < L$.
On the other hand, any $P_L^\pi$ can be obtained through depolymerisation of 
one of $2n$ types of polymers of length $L+1$, one of which being  $P_{L+1}^\pi$ 
and the remaining $2n-1$ being of type $P_{L+1}^\al$  with $\alpha\{L+1\}=(\pi\{L\},\alpha_{L+1}),\, \alpha_{L+1}\neq \pi_{L+1}$, or $\alpha\{L+1\}=(\alpha_1,\,\pi\{L\})$ with $\alpha_1$ any of the $n$ types of amino acids. More generally, they can be obtained from $P_{L+\ell^\prime}^\pi$ and 
$2n^{\ell^\prime}-1$ polymers of type $P_{L+\ell^\prime}^\al$ where $\ell^\prime  \ge 1$ and $\alpha\{L+\ell^\prime\}=(\pi\{L\},\,\alpha_{L+1},\ldots \alpha_{L+\ell^\prime})$ with $\alpha_j \neq \pi_j, j=L+1,\ldots L+\ell^\prime$, or $\alpha\{L+\ell^\prime\}=(\alpha_1,\ldots \alpha_{\ell^\prime},\,\pi\{L\})$ for any type $\alpha_j, j=1,\ldots \ell^\prime$. 
The same is true for the corresponding polynucleotide chains.

When the polymerase is bound to a polynucleotide, it  becomes more 
stable either through induced folding of a (partially) unfolded sequence, or 
through the inaccessibility of  bound portions, or both.
We thus 
define $F_\pi (\ell)$ as the depolymerisation reduction coefficient for the 
bound polymerase of length $\ell$, with that reduction coefficient being 1 when no depolymerisation occurs at all. We estimate it to be 
\begin{eqnarray*}
F_\pi (\ell) = \left\{\begin{array}{ll}
 1- e^{\displaystyle{-\frac{\ell-\ell_{\pi {\rm min}}+1}{\lambda}}} & \ell \,\ge\, \ell_{\pi {\rm min}}\\
0 & \ell\, <\, \ell_{\pi {\rm min}}
\end{array}\right .
\end{eqnarray*}
with $\lambda\,> 0$ a parameter controlling how much of the polymerase is 
stabilised. The term ${(\ell-\ell_{\pi {\rm min}}+1)}/{\lambda}$ can be 
interpreted as a Boltzmann factor with a free energy expressed in units of $k_BT$.
The hydrogen bond binding energy between RNA and a polypeptide is 
approximately $16 {\rm kJ/}{\rm mol}$ [\cite{Dixit:2000}], so  assuming that 
the number of such hydrogen bonds between the polymerase and the polynucleotide
is $\ell-\ell_{\pi {\rm min}}+1$, one has $\lambda\approx 0.15$.

The binding rate of a polymerase to a polynucleotide $R_M^\alpha$ of 
length $M$ and sequence $\alpha$ is 
$k_{b,M}=Z\,R_M^{\alpha}\,n^M$ where  $n^M$ is the total number of polynucleotides of length $M$.  The probability that a polymerase of 
length $L$ binds to a polynucleotide of length $M$ is therefore given by 
\begin{eqnarray*}
\widetilde{{\cal P}}_{b,M}&=&\frac{k_{b,M}}
  {\sum_{m=2}^{L_{\rm max}} k_{b,m}}.
\end{eqnarray*}
The total time the polymerase remains bound to a  polynucleotide of 
length $M$ is  estimated  to be $M/k_{R+}$. Therefore  the probability ${\cal P}_b$ for
a polymerase  to be bound is given by the average binding time 
divided by the sum of the average binding time and the average time needed to 
bind:
 \begin{eqnarray*}
 {\cal P}_b=\frac{\sum_{M=2}^{L_{\rm max}}\frac{M}{k_{R+}} \widetilde{{\cal P}}_{b,M}}
  {\sum_{M=2}^{L_{\rm max}}\left(\frac{M}{k_{R+}}\widetilde{{\cal P}}_{b,M}\right) 
     + \displaystyle{\frac{1}{\sum_{m=2}^{L_{\rm max}} k_{b,m}}}}.
\end{eqnarray*}
As a result the polymerase depolymerisation rate will  be
\begin{eqnarray*}
K_{P,\al,L}^{-} &=& K_{P}^{-}, \nonumber\\
K_{P,\bpi,L}^{-} &=& K_{P}^{-}, \nonumber\\
K_{P,\pi,L}^{-} &=& K_{P}^{-} (1-{\cal P}_b F_{\pi}(L)).
\end{eqnarray*}

\item[{\em (d)}] {\em  Equations}

For any chain of length $\ell$, our model considers the 
concentrations of  polynucleotides and polypeptides corresponding to the 
polymerase sequence $\pi$, its complementary sequence $\bpi$ and  the generic
sequences  $\alpha$. We assume that the concentrations of  polynucleotides
and polypeptides of a specific length, bar the polymerase and its 
complementary sequence, are identical. For
the chains that share the first $\ell$ elements  of their sequence with those of 
the polymerase (or its complementary chain),  and differ in all other elements,
this is only an approximation, but it
is nevertheless justified, as the concentrations 
of these polymers  only differ slightly from those of polymers with sequences of type
$\al$, and
their contribution to the variation of the 
polymerase concentration is expected to be small.

The variations in polymer concentrations as time evolves are governed in 
our model by a system of ordinary differential equations.
In the equations, $L$ is the length of the polymer chains, spanning all 
values in the range $1< L\le L_{\rm max}$ where $L_{\rm max}$ is the maximal length 
of polypeptide and polynucleotide chains. We thus have a system of 
 $6\times (L_{\rm max}-1)$ equations.  We recall that $n$ is the number of codon 
types, assumed to be equal to the number of amino acid types. 
 
\begin{eqnarray}
\frac{d R_L^{\pi}}{dt} &=& 
K_R^+ R_{1}R_{L-1}^{\pi}-nK_R^+R_{1}R_{L}^{\pi}
+ \sum_{\ell=L+1}^{L_{{\rm max}}}\left[K_R^- R_{\ell}^{\pi} +(2n^{\ell-L}-1) K_R^- R_{\ell}^{\al}\right] 
\nonumber\\
&&
-(L-1)K_R^-R_{L}^{\pi} +{\cal K}(L) R_L^{\bpi}
\nonumber\\
\frac{d R_L^{\bpi}}{dt} &=& 
K_R^+ R_{1}R_{L-1}^{\bpi}-nK_R^+R_{1}R_{L}^{\bpi}
+ \sum_{\ell=L+1}^{L_{{\rm max}}}\left[K_R^- R_{\ell}^{\bpi} +(2n^{\ell-L}-1) K_R^- R_{\ell}^{\al}\right]
\nonumber\\
&&
-(L-1)K_R^-R_{L}^{\bpi}+{\cal K}(L)R_L^{\pi}
\nonumber\\
\frac{d R_L^{\al}}{dt} &=& 
K_R^+R_{1}R_{L-1}^{\alpha}-nK_R^+R_{1}R_{L}^{\al}
+ 2\sum_{\ell=L+1}^{L_{{\rm max}}} n^{\ell-L} K_{R}^- R_{\ell}^{\al}-(L-1)K_R^-R_{L}^{\al}
\nonumber\\
&&
+{\cal K}(L)R_L^{\al}
\nonumber\\
\frac{d P_L^{\pi}}{dt} &=& 
K_P^+P_1P_{L-1}^{\pi}-nK_P^+P_1P_{L}^{\pi}
+ \sum_{\ell=L+1}^{L_{{\rm max}}} \left[K_{P}^-(1-{\cal P}_b\,F_{\pi}(L)) P_{\ell}^{\pi}+ (2n^{\ell-L}-1) K_{P}^- P_{\ell}^{\al}\right]
\nonumber\\
&&
-(L-1)K_{P}^-(1-{\cal P}_b\,F_{\pi}(L)P_{L}^{\pi}+k_{P,L}^+ P_1R_L^{\pi}
\nonumber\\
\frac{d P_L^{\bpi}}{dt} &=& 
K_P^+P_1P_{L-1}^{\bpi}-nK_P^+P_1P_{L}^{\bpi}
+\sum_{\ell=L+1}^{L_{{\rm max}}} \left[K_{P}^- P_{\ell}^{\bpi}+  (2n^{\ell-L}-1) K_{P}^- P_{\ell}^{\al}\right]
\nonumber\\
&&
-(L-1)K_{P}^-P_{L}^{\bpi}+k_{P,L}^+ P_1R_L^{\bpi}
\nonumber\\
\frac{d P_L^{\al}}{dt} &=& 
K_P^+P_1P_{L-1}^{\alpha}-nK_P^+P_1P_{L}^{\al}
+ 2\sum_{\ell=L+1}^{L_{{\rm max}}} n^{\ell-L} K_{P}^- P_{\ell}^{\al}-(L-1)K_{P}^-P_{L}^{\al}
\nonumber\\
&&
+k_{P,L}^+ P_1R_L^{\al}.
\label{eq_EQL}
\end{eqnarray}

Alongside the seven physical parameters $\{K^\pm_R,\, K^\pm_P\,, h^+_{P,L},\, {\cal K}(L),\,{\cal P}_b\}$ appearing in the differential equations above,  we need to consider two parameters yielding the  `initial' concentrations of amino acid and nucleotide inside the system, namely  $\rho_p\equiv P_1(t=0)$ and $\rho_r\equiv R_1(t=0)$.
In the absence of actual data for these quantities, we explore a range of realistic values in the analysis of our model.
The concentration of free amino acids and nucleotides at any one time
is then given by 
$P_1(t)=\rho_p -\sum_{L=2}^{L_{{\rm max}}} [(n^L-2) P_L^\al(t) +P_L^\pi(t)+P_L^\bpi(t) ]$ and $R_1(t)=\rho_r -\sum_{L=2}^{L_{{\rm max}}}[ (n^L-2) R_L^\al(t)+R_L^\pi(t)+R_L^\bpi(t)]$ respectively, with $P_L^S(0)=R_L^S(0)=0$ for any value of $L$ in the range $2 \le L \le L_{\rm max}$ and sequence $S=\alpha, \pi, \bar{\pi}$.

\end{enumerate}

{\em Results}

The system of equations [Equation \eqref{eq_EQL}] is non-linear and  too 
complex to solve analytically. We therefore analyse it numerically,  
starting from a system
made entirely of free nucleotides, amino acids, 
as well as  charged p-tRNA, and letting the system 
evolve until it settles into a steady configuration. 

The main quantities of interest are  the relative concentrations of the polymerase ($\rho_{\pi}$) and of
 the $\alpha$ peptide chains ($\rho_{\alpha}$). We have
\begin{equation*}
\rho_\pi = \sum_{\ell=\ell_{\pi {\rm min}}}^{L_{{\rm max}}} P_\ell^\pi\qquad {\rm and}\qquad 
\rho_\alpha = \sum_{\ell=\ell_{\pi {\rm min}}}^{L_{ {\rm max}}} P_\ell^\alpha,
\end{equation*} 
and  evaluate the ratios
\begin{equation*}\label{ratios}
Q_1 = \frac{\rho_\pi}{\rho_\alpha} \qquad {\rm and}\qquad 
Q_{2,\ell} =\frac{P_{\ell}^\pi}{P_{\ell}^\alpha}
\end{equation*}
while monitoring the evolution of each quantity over time. 
$Q_1$ corresponds to the relative amount of polymerase of any length compared
to other proteins (for a specific arbitrary sequence $\al$), while $Q_{2,\ell}$ corresponds
to the relative amount of polymerase of length $\ell$ compared to an arbitrary 
protein of length $\ell$. Unit ratios indicate that the polymerase has not
been selected at all, while large values of $Q_1$ or $Q_{2,\ell}$ on the other hand
indicate a good selection of the polymerase. 

The complexity of the system [Equation \eqref{eq_EQL}] also lies in the 
number of free parameters it involves. A systematic analysis of the high-dimensional parameter space 
is beyond the scope of this article, and we therefore concentrate on the analysis and description of results for a selection of parameter values that highlight potentially interesting behaviours of our model.

Recall that our model assumes that the number $n$ of different amino acids is equal to the number of codon types, and throughout our numerical work we have set $n=4$. Note that the word `codon' here is used by extension. Indeed, there are four different nucleic acids in our model and  the `biological' codons are made of two nucleic acids, bringing their number to sixteen. However, they split into four groups of four, each of which encoding one of the four amino acids.
From a mathematical modelling point of view, this is completely equivalent.
 It is well accepted that early proteins were produced using a reduced set of amino acids (\citealt{Angyanetal:2014}). The exact identity and number is unclear though experimental work has shown that protein domains can be made using predominantly five amino acids (\citealt{Riddle:1997}) while the helices of a 4-alpha helix bundle were made using only 4 amino acids 
(\citealt{ReganDeGrado:1988}).
We have used mostly 
$\ell_{\pi {\rm min}}=7$ and $\ell_{\pi {\rm max}}= L_{\rm max}=10$, but have  
investigated other values as well.

While these figures are  somewhat arbitrary, an $\ell_{\pi {\rm min}}$ of 7 was chosen as it corresponds to the typical minimum number of amino acids required to produce a stable, folded alpha helix structure (\citealt{Manning:1988}). The choice of $L_{\rm max}=10$ is based on the fact that while the polymer peptide chains could be significantly longer, they would need correspondingly long polynucleotide sequences to encode them, which becomes increasingly unlikely as lengths increase.
Furthermore, we expected polymers of length 10 to have very low concentrations, a hypothesis confirmed by our simulations. We have nevertheless investigated larger values of $L_{\rm max}$ as well, and found little difference, as outlined below.

In a first step, guided by data on parameter values gleaned from the literature and gathered in the Supplementary Information section, we set
\begin{eqnarray}
\begin{array}{ll}
K_R^+=4.2 \times 10^{-7}\, {\rm mol}^{-1}\,{\rm m}^3\, {\rm s}^{-1},&
K_R^{-}= 8 \times 10^{-9} {\rm s}^{-1},\\
K_P^+= 2.8\times 10^{-21}\,{\rm mol}^{-1}\,{\rm m}^3 \, {\rm s}^{-1},&
K_{P}^-= 4 \times 10^{-11}{\rm s}^{-1}\\
k_{P,1}^+= 0.1\, {\rm mol}^{-1}\,{\rm m}^3\, {\rm s}^{-1},&
h_{R} = 10^6\,{\rm mol}^{-1}\,{\rm m}^3\, {\rm s}^{-1},\\
Z= 10^6\, {\rm mol}^{-1}\,{\rm m}^3\, {\rm s}^{-1},&
\lambda = 0.15,\\
k_{{\rm step}}=4\times\,10^{-5}{\rm s}^{-1}.&
\end{array}
\label{eq_pars}
\end{eqnarray}
We let the system evolve under a variety of
initial concentrations of free amino acids and nucleotides, 
$\rho_p$ and $\rho_r$,  in the range 
$10^{-5} - 0.1\, \mbox{mol\,m}^{-3}$, and with all polymer concentrations set to $0$. 
We monitored the concentration of all  
polymers, in particular the concentration of polymerase $\rho_\pi$ and its ratio to the concentration of $\alpha$ polypeptide chains, $Q_1$. In most cases we found 
that the nucleotides polymerised spontaneously (Mechanism (A)) in small amount and this led, 
indirectly, to the polymerisation of the polypeptides, including the 
polymerases (Mechanism (C)). The polymerases then induced further polymerisation of the 
polynucleotides (Mechanism (B)) and the system slowly equilibrated. 

\begin{figure}[!ht]
\begin{tabular}{ll}
\includegraphics[width=7.5cm, keepaspectratio]{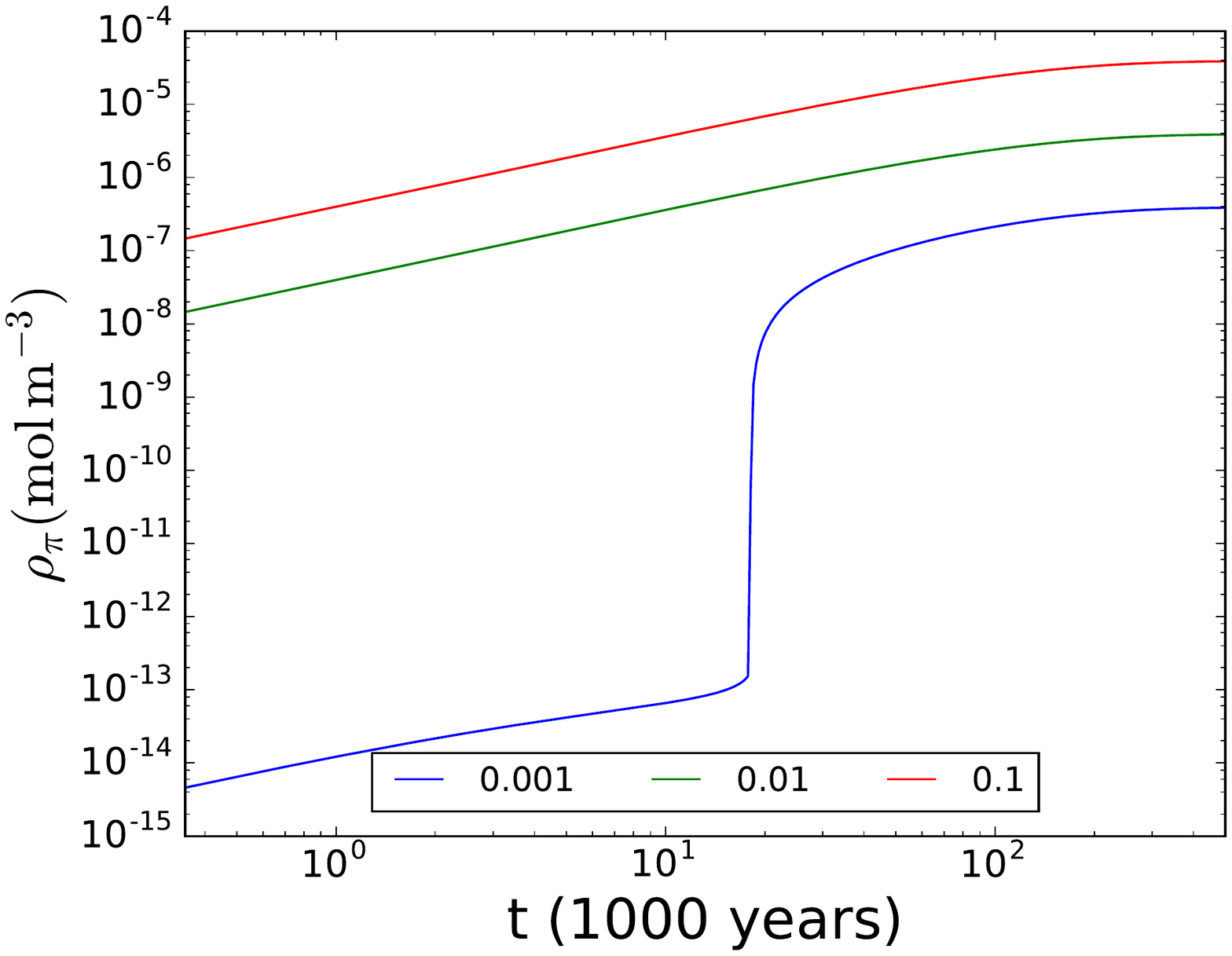}
&\includegraphics[width=7.5cm, keepaspectratio]{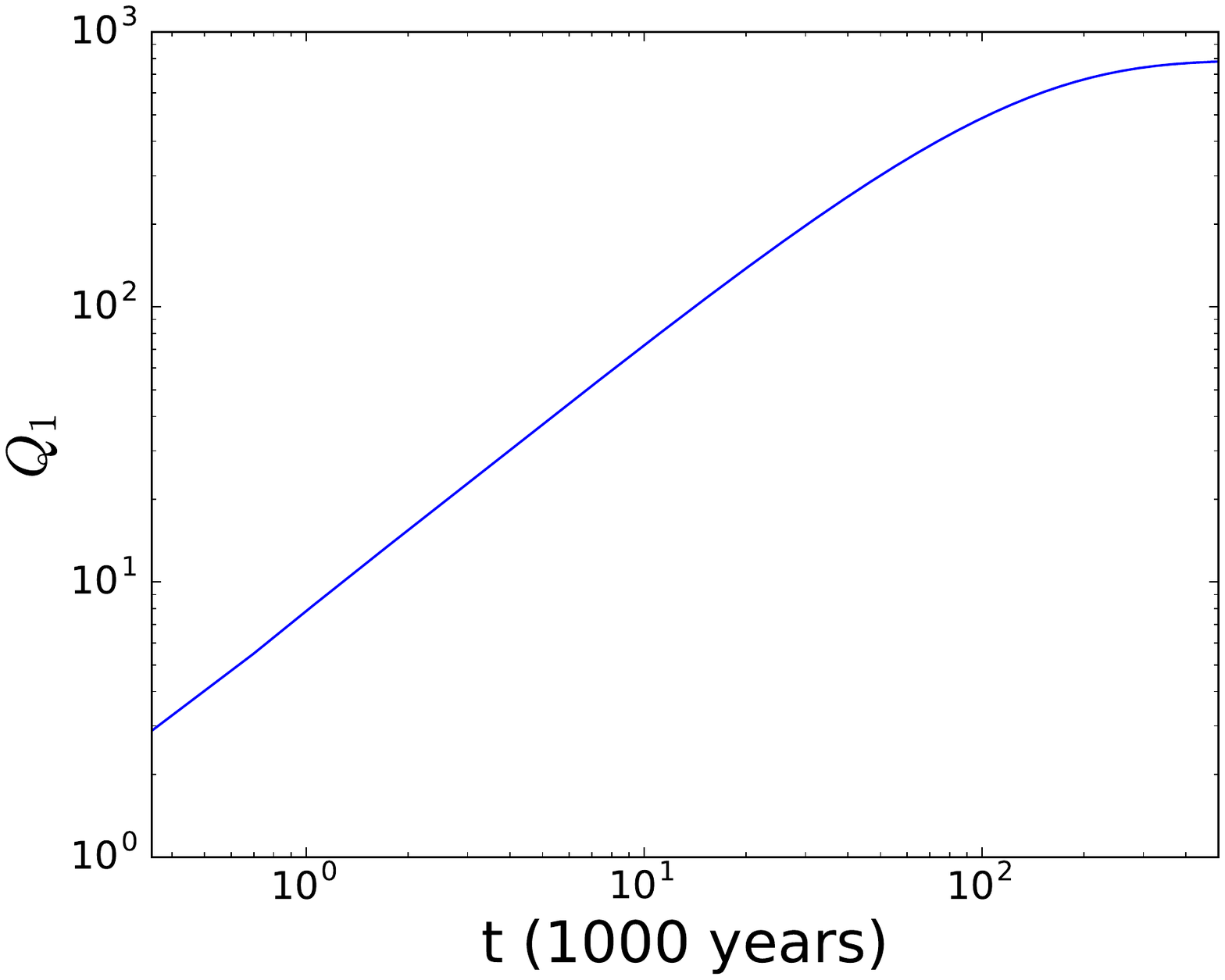}
\\
\vspace{-30mm}
\hbox to 75mm{\hspace{30mm} (a)\hspace{30mm}} & 
\hbox to 75mm{\hspace{30mm} (b)\hspace{30mm}}\\
\end{tabular}
\caption{\label{fig_Q1}
a) Time evolution of the polymerase for initial concentration 
$\rho_r=\rho_p=0.001$, $0.01$ and $0.1 \,{\rm mol}\,{\rm m}^{-3}$.
b) $Q_1$ for initial concentration 
$\rho_r=\rho_r=0.01\,{\rm mol}{\rm m}^{-3}$.
Parameter values: $K_P^-=4 \times 10^{-11} {\rm s}^{-1}$, $Z=10^6\,{\rm mol}^{-1}\,{\rm m}^3\,{\rm s}^{-1}$, 
$\lambda=0.15$.
}
\end{figure}

The end result was an excess of polymerase of all lengths compared to 
$\alpha$ polypeptide chains with $Q_1=786$  for all 
initial concentrations $\rho_p=\rho_r \ge 0.001\, \mbox{mol\,m}^{-3}$
(Figure \ref{fig_Q1}).
Moreover the total amount of polymerase reached, for initial concentration 
of free amino acids $\rho_p$, was 
a concentration of approximately $4\times\,10^{-4}\times\,\rho_p$ (as 
illustrated by the bottom 2 rows in Table 1). 
The concentration of polymerase of length 10, on the other hand, was 
very small $P^\pi_{10}=6.3\times 10^{-14}{\rm mol}\,{\rm m}^{-3}$ for 
but $Q_{2,10}=5.9\times 10^{18}$ was very large, effectively showing that the only 
polypeptide chain of length $L_{\rm max}= 10$ was the polymerase. 

We  found hardly any  polymerisation of the polymerase 
when $\rho_p=\rho_r = 0.0009\, \mbox{mol\,m}^{-3}$, with 
$\rho_\pi\approx 1.4\times 10^{-14}\,{\rm mol}\,{\rm m}^{-3}$ and $Q_1=12.4$, 
while with $\rho_p=\rho_r = 0.001\, \mbox{mol\,m}^{-3}$, we obtained
$\rho_\pi\approx 3.9\times 10^{-7}\,{\rm mol}\,{\rm m}^{-3}$ and  
$Q_1=786$ 
(fig \ref{fig_Q1} a). This highlights  a very sharp transition at a {\em critical concentration} $\rho_{p,c}$ above which polymerases are generated. 
We summarise the data in Table \ref{table:transition}.
\begin{table}[h]
\centering
\begin{tabular}{c|c|c|c|c}
$\rho_{p}\, ({\rm mol}\,{\rm m}^{-3})$&$\rho_{r}\,({\rm mol}\,{\rm m}^{-3})$ &$\rho_\pi\,\,({\rm mol}\,{\rm m}^{-3})$&$Q_1$&polymerase production\\[-8pt]
&&&&\\
\hline
&&&&\\[-8pt]
$2 \times 10^{-4}$&$2 \times 10^{-4}$&$2.8\,10^{-19}$&$1.0008$&insignificant \\
$9 \times 10^{-4}$&$9 \times 10^{-4}$&$1.4\,10^{-14}$&$12.4$&insignificant \\
$10^{-3}$         &$10^{-3}$         &$3.9\times 10^{-7}$&$786$&yes\\
$10^{-1}$         &$10^{-1}$         &$3.9\times 10^{-5}$&$786$&yes\\
\end{tabular}
\caption{Sharp transition in the production of polymerases due to variations in the initial concentrations of free peptides and nucleotides, all other parameters kept fixed at the values [Equation \eqref{eq_pars}]. }\label{table:transition}
\end{table}

We then fixed the initial concentration $\rho_p$ to four different values 
and varied $\rho_r$ to identify the critical initial concentration of 
nucleotides necessary for the production of polymerases. The results in 
Table \ref{table:criticalrho} show that the critical concentration 
$\rho_{r,c}$ is nearly constant and of the order of 
$10^{-3}\,{\rm mol}\,{\rm m}^{-3}$ for a very wide
range of amino acid initial concentrations.  
\begin{table}[h]
\centering
\begin{tabular}{c|c}
$\rho_{p}\, ({\rm mol}\,{\rm m}^{-3})$&$\rho_{r,c}\,({\rm mol}\,{\rm m}^{-3})$ \\[-8pt]
&\\
\hline
&\\[-8pt]
$10^{-4}$&$2 \times 10^{-3}$ \\
$10^{-3}$&$10^{-3}$\\
$10^{-2}$&$8\times10^{-4}$\\
$10^{-1}$&$7\times10^{-4}$\\
\end{tabular}
\caption{Values of the free nucleotide initial critical concentration given the initial concentration of free peptides. The other parameters are kept fixed at the values [Equation \eqref{eq_pars}]. }\label{table:criticalrho}
\end{table}

Many of the parameters we have used were estimated or measured in 
conditions which, in all likelihood, were not identical to the ones existing
when the 
polymerisation we are modelling occurred. In a second step, we departed from
the set of values [Equation \eqref{eq_pars}] and found that in all cases investigated, 
varying these parameters modified the critical concentrations of 
$\rho_{r,c}$ and $\rho_{p,c}$, but did not affect significantly the
value of $Q_1$ while $Q_{2,10}$ remained extremely large. 

More specifically, 
taking 
$K_P^-=5.1\times 10^{-6}{\rm s}^{-1}$ marginally increased the critical 
concentration to $\rho_{r,c}=\rho_{p,c}=0.0011\,{\rm mol}\,{\rm m}^{-3}$.
Similarly, taking $k_{{\rm step}}=0.02{\rm s}^{-1}$ 
increased slightly the critical concentrations: 
$\rho_{r,c}=\rho_{p,c}=0.0017\,{\rm mol}\,{\rm m}^{-3}$. 
On the other hand, taking
$Z=10^8{\rm mol}^{-1}\, {\rm m}^3\,{\rm s}^{-1}$ lead to a decrease of the 
critical concentrations: $\rho_{r,c}=\rho_{p,c}=0.0005\,{\rm mol}\,{\rm m}^{-3}$.
Varying $h_R$ to values as small as $1\,{\rm mol}^{-1} {\rm m}^3 {\rm s}^{-1}$ did not
change the critical concentrations.

In our model we have considered  the concentrations of free amino acids ($\rho_p\equiv P_1$)
and charged p-tRNA to be identical: $k_t\approx 1$ (see [Equation \eqref{ptRNAP1}].  To consider 
other values of $k_t$ we only need to multiply the polymerisation rate of a peptide ($k_{P,1}^+$) by $k_t$
as it is p-tRNAs that bind to XNA chains, not free amino acids.
We have  considered a large range of values for $k_{P,1}^+$ and found that 
for $k_{P,1}^+=10^{-5}\, {\rm mol}^{-1}\,{\rm m}^3\, {\rm s}^{-1}$, the critical 
concentrations had not changed significantly while for
$10^{-8}\,{\rm mol}^{-1}\,{\rm m}^3 \, {\rm s}^{-1}$, they increased to 
$\rho_{r,c}=\rho_{p,c}=0.002\,{\rm mol}\,{\rm m}^{-3}$. This shows that taking 
much smaller values of $k_t$ has a very small impact on our results and 
that having a concentration of charged p-tRNA much smaller than that of free 
amino acids would only increase marginally the critical concentrations we have 
obtained using our original assumption.


The parameters on which the model is the most sensitive are $K_R^+$ and 
$K_R^-$.
We found that for $K_R^+=4\times 10^{-8}\,{\rm mol}^{-1}\,{\rm m}^3 \, {\rm s}^{-1}$,
$\rho_{r,c}=\rho_{p,c}=0.007\,{\rm mol}\,{\rm m}^{-3}$
and for $K_R^+=4\times 10^{-9}\, {\rm mol}^{-1}\,{\rm m}^3\,{\rm s}^{-1}$,
$\rho_{r,c}=\rho_{p,c}=0.05\,{\rm mol}\,{\rm m}^{-3}$. 
Similarly, for $K_R^-=10^{-7}\,{\rm s}^{-1}$ we found that 
$\rho_{r,c}=\rho_{p,c}=0.01\,{\rm mol}\,{\rm m}^{-3}$
and for $K_R^-=10^{-6}\,{\rm s}^{-1}$ that 
$\rho_{r,c}=\rho_{p,c}\approx 0.19\,{\rm mol}\,{\rm m}^{-3}$. 
This shows that the spontaneous polymerisation of polynucleotide is 
essential 
to reach a minimum concentration of polynucleotides to kick start the whole 
catalysis process and  that the stability of the polynucleotides 
plays an important role.

To investigated this, we have run simulations with 
$K_R^+=4\times 10^{-8}\, {\rm mol}^{-1}\,{\rm m}^3\, {\rm s}^{-1}$ for a fixed duration, 
$\tau_{\rm pol}$, after which $K_R^+$ was set  to $0$. We found that if 
$\tau_{\rm pol}$ was long enough, the polymerisation of polypeptide and 
polynucleotide chains was identical to the one obtained when $K_R^+$ was not
modified. When $\tau_{\rm pol}$ was too short, on the other hand, one was only left 
 with short polypeptide and polynucleotide chains in an equilibrium 
controlled by the spontaneous polymerisation and depolymerisation parameters.
The minimum value for $\tau_{\rm pol}$ depends on
the concentrations $\rho_{r}$ and $\rho_{p}$ and the results are given in Table \ref{table:one}.
\begin{table}[h]
\centering
\begin{tabular}{l||c|c|c|c}
$\rho_{r}=\rho_{p}\, \,({\rm mol}\,{\rm m}^{-3})$ &$0.001$&$0.002$&$0.005$&$0.01$\\[-8pt]
&&&&\\
\hline
&&&&\\[-8pt]
$\tau_{\rm pol}\,\,(\rm years)$ &18000 & 254 & 12.7 years & 2.2 \\
\end{tabular}
\caption{
$\tau_{pol}$ is the minimum duration of spontaneous polynucleotide polymerisation,
given here for different initial concentrations of free nucleotides $\rho_{r} (=\rho_{p})$ 
needed to induce large polymerase concentrations.} 
\label{table:one}
\end{table}

This shows that while $K_R^+$ is an important parameter in the process, what 
matters is to have a spontaneous generation of polynucleotides at the onset (Mechanism (A)). 
This then leads to the production of polypeptides, including polymerase (Mechanism (C)) and, 
once the concentration of polymerase is large enough, the catalysed 
production of polynucleotides (Mechanism (B)) dominates the spontaneous polymerisation.

We have also varied $K_R^-$ once the system had settled and we found that 
for $\rho_{r}=\rho_{p}=0.01 \,{\rm mol}\,{\rm m}^{-3}$, $K_R^-$ could be increased
up to $6\times 10^{-7}{\rm s}^{-1}$ while still keeping a large amount of polymerase.
Above that value, the polynucleotides are too unstable and one ends up
again with mostly short polymer chains and $Q_1\approx 1$. 

We have also considered values of $L_{\rm max} > 10$ and found that the main 
difference is a slight increase of the critical concentrations. For example, 
for $L_{\rm max} =11, 12$ and $15$, $\rho_{r,c}=\rho_{p,c}$ are respectively 
equal to $0.001, 0.0011$ and $0.0011 \,{\rm mol}\,{\rm m}^{-3}$. At given 
concentrations $Q_1$ and $\rho_\pi$ remain unchanged but $P_{L_{\rm max}}^\pi$ 
deceases approximatekly by a factor of $40$ each time $L_{\rm max}$ is increased 
by 1 unit.

We have also taken $L_{\pi {\rm min}}= 4,5$ and $6$ and found that the  critical 
concentrations were respectively $2\times 10^{-5}, 2\times 10^{-4}$ and 
$4\times 10^{-4}  \,{\rm mol}\,{\rm m}^{-3}$, while $\rho_\pi$ took the values
of approximately $0.01$, $2.5\times 10^{-3}$ and $3\times 10^{-4} \,{\rm mol}\,{\rm m}^{-3}$. $Q_1$ on the other hand remained constant.

A summary of the parameter values investigated outside the set [Equation \eqref{eq_pars}] and the corresponding critical concentrations are given in Table \ref{table:summary}. Only one parameter was changed at a time.
\begin{table}[h]
\centering
\begin{tabular}{c|l}
\sc{Modified Parameter} & $\rho_{r,c}=\rho_{p,c}\,\, (\rm {mol\,m}^{-3})$\\[5pt]
\hline
&\\[-8pt]
$K_P^-=5.1 \times 10^{-6}\, {\rm s}^{-1}$& $1.1\times 10^{-3}$\\[2pt]
$k_{{\rm step}}=2\times 10^{-2}\, {\rm s}^{-1}$ & $1.7\times 10^{-3}$\\[2pt]
$Z=10^8\, {\rm mol}^{-1}\,{\rm m}^{-3}{\rm s}^{-1}$ & $5\times 10^{-4}$\\[2pt]
$h_R=1\, {\rm mol}^{-1}\,{\rm m}^{-3}{\rm s}^{-1}$&$10^{-3}$ \\[2pt]
$k_{P,1}^+=10^{-5}\, {\rm mol}^{-1}\,{\rm m}^3\,{\rm s}^{-1}$&$10^{-3}$\\[5pt]
$k_{P,1}^+=10^{-8}\, {\rm mol}^{-1}\,{\rm m}^3\,{\rm s}^{-1}$&$2\times 10^{-3}$\\[2pt]
$L_{\rm max}=15$  & $1.1\times 10^{-3}$\\[2pt]
$L_{\pi {\rm min}}=6$& $4\times 10^{-4}$\\[2pt]
$L_{\pi {\rm min}}=5$& $2\times 10^{-4}$\\[2pt]
$L_{\pi {\rm min}}=4$& $2\times 10^{-5}$\\[2pt]
\hline
&\\[-8pt]
$K_R^+=4\times 10^{-8}\, {\rm mol}^{-1}\,{\rm m}^3\,{\rm s}^{-1}$&
$7\times 10^{-3}$\\[2pt]
$K_R^+=4\times 10^{-9}\, {\rm mol}^{-1}\,{\rm m}^3\,{\rm s}^{-1}$&
$5\times 10^{-2}$\\[2pt]
$K_R^-=10^{-7}\,{\rm s}^{-1}$& $10^{-2}$ \\[2pt]
$K_R^-=10^{-6}\,{\rm s}^{-1}$& $0.19$ \\[2pt]
\end{tabular}
\caption{ Free nucleotide and peptide initial critical concentrations for a set of parameters differing from  [Equation \eqref{eq_pars}] in one parameter at a time (listed as `modified parameter' in the table). For reference, $\rho_{r,c}=\rho_{p,c}=10^{-3}(\rm {mol\,m}^{-3})$ for the set [Equation \eqref{eq_pars}].}\label{table:summary}
\end{table}

%

{\bf Discussion}

We describe a theoretical nucleopeptidic reciprocal replicator comprising a polynucleotide that templates the assembly of small p-tRNA adapter molecules, most likely having mixed backbone architectures. These spontaneously arising p-tRNAs would have been bound to various classes of amino acids (possibly via weak stereochemical specificity), and a simple increase in local concentration mediated by binding to the p-Rib (in its most primitive version nothing much more than a mixed backbone architecture p-mRNA) could have driven polypeptide polymerisation. Once a template arose that coded for a peptide able to catalyze phosphodiester bond formation, this p-Rib could have templated assembly of its own complementary strand (and vice versa) and the self-replication cycle would have been complete. (See Figure 6 for a summary).

In reality, the IDA may have comprised a distribution of related sequences of peptides and XNAs. We can imagine that over time, different p-Ribs encoding different peptides with additional functionalities could have appeared as the system evolved and that these p-Ribs may have subsequently fused together into larger molecules. 


By imagining the IDA as a pool of molecules where variety within the ``species'' is maintained by the poor copying fidelity of a statistical operational code, should any mutation that stops replication arise, the other molecules in the pool would still function, ensuring continuity of the whole. Indeed this could have provided a selective pressure for superior replicators. While our model does not directly consider less than perfect copying fidelity, it is not expected to have a major effect on our conclusions as copies with decreased performance would not be maintained as a significant proportion of the population and copies with increased performance would simply take over the role of main replicator.

\begin{figure}[ht!]
\begin{center}
\includegraphics[width=12.5cm, keepaspectratio]{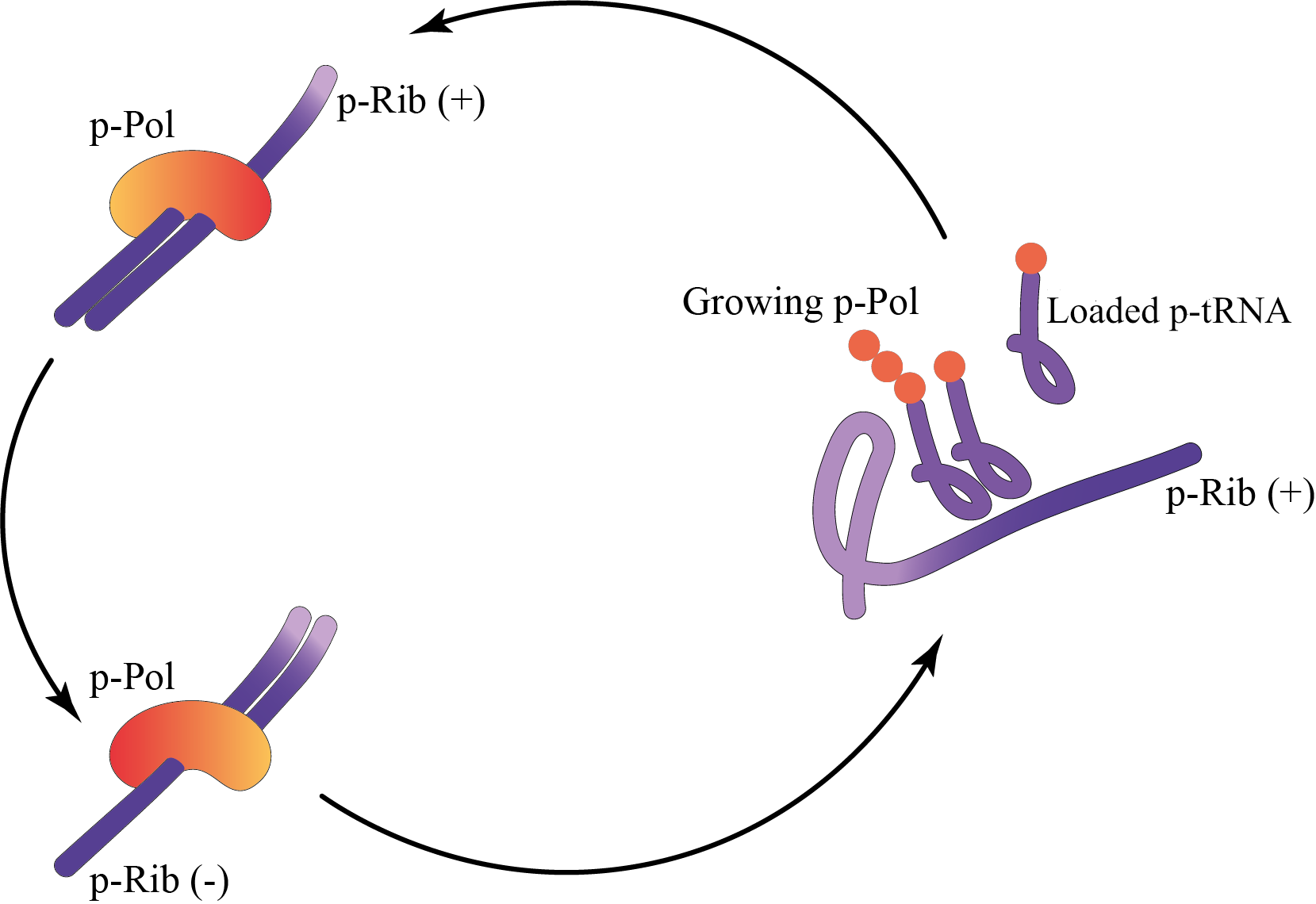}
\caption{\label{fig6} The nucleopeptide Initial Darwinian Ancestor. In this cartoon model a short strand of XNA has the functionality of both a primordial p-mRNA and a p-Rib. Primitive XNA molecules loaded with amino acids (p-tRNA) bind to the p-Rib via codon-anticodon pairing. This allows adjacent amino acids to undergo peptide bond formation and a short peptide chain is produced. A certain peptide sequence is able to act as a primordial XNA-dependent XNA polymerase (p-Pol) able to copy both + and $-$ p-Rib strands to eventually produce a copy of the p-Rib(+) strand. }
\end{center}
\end{figure}

The primordial operational code may only have required two bases per p-tRNA to deliver statistical proteins, while the catalytic requirements of the p-Pol are loose enough that a 7-residue peptide is a plausible lower length limit. This reduces the minimum length of the posited spontaneously arising p-Rib to just \b{14 nucleotides (assuming no spaces between codons)}. This is an optimistic length estimate, but given the available time and with molecular co-evolution, inorganic catalysts and geological PCR, considerably longer molecules may have been possible (\citealt{Baaskeetal:2007, Fishkis:2010}). These p-Pols would act on  p-Ribs and the crucial abiogenesis step would be the emergence of a 14-mer XNA that, in the context of the primordial operational code, happened to code for a peptide able to bind XNA and catalyze phosphodiester bond formation of base-paired nucleotides. Although the concentrations of various components are not known with certainty this does not seem an unreasonable proposition particularly given that functional peptides are known to occur in random sequences with surprising frequency (\citealt{KeefeSzostak:2001}). 

Our mathematical model showed that the most important parameters, apart 
from the concentration of loaded p-tRNA and polynucleotides, are the spontaneous
polymerisation and depolymerisation of polynucleotides. It also shows that 
polynucleotides are first polymerised spontaneously and that these initial
polynucleotides catalyse the production of the first polypeptides, including the polymerase. 
These polymerases can then generate further polynucleotides through catalysis.
The stability gained by polymerases while being bound to polynucleotides 
ultimately leads to an increase of their relative concentration compared to 
the other polypeptides.

Overall, the hypothesis explains the coupling of polynucleotide and polypeptide polymerisation, the operational code and mutations in the p-Pol sequence that could eventually result in increased specificities leading to primitive DNA polymerases and RNA polymerases. No extraordinary exchanges of function are required and each molecule is functionally similar to its present-day analogue. Like all new abiogenesis theories, this IDA requires in vitro confirmation; in particular, the steps required for the primordial operational code to arise {\it ab initio} warrant close attention. 

The idea that the ancestral replicator may have consisted of both nucleic acid and peptide components (the `nucleopeptide world') is in itself not new, but  compared to the RNA world, has been somewhat neglected. We argue that molecular co-evolution of polynucleotides and peptides seems likely and cross-catalysis is known to be possible, for example in vitro selection experiments delivered RNA with peptidyl transferase activity after just nine rounds of a single selection experiments (\citealt{ZhangCech:1997, Fishkis:2010}). Inversely, Levy and Ellington produced a 17-residue peptide that ligates a 35 base RNA (\citealt{LevyEllington:2003}).

Nucleopeptide world research is relatively sparse, the data collected so far hint that cross-catalysis may be more efficient than autocatalysis by either peptides or nucleic acids. A self-replicating primordial system wherein RNA encoding for protein was replicated by a primordial RNA-dependent RNA polymerase which carried out the role of a replicative agent rather than as a transcriber of genes has previously been suggested (\citealt{Leipeetal:1999}), although in this case no further development of the concept to produce a self-contained replicating system was pursued. The merits of a ``two polymerase'' system where RNA catalyses peptide polymerisation and vice versa were succinctly explained by Kunin (\citealt{Kunin:2000}), although possible mechanisms and validity were not considered in detail. Our IDA hypothesis has tried to set out more rigorously the possible steps and processes whereby a nucleopeptide IDA could have arisen and could be tested experimentally.

Future experimental work that would support the nucleopeptide theory would be to provide evidence that the stereochemical hypothesis applies to the earliest occurring amino acids including those likely to have composed the active site of the p-Pol. Currently codon/anticodon binding to a number of amino acids has been shown (\citealt{Yarusetal:2005b}) but is absent for the four earliest amino acids (\citealt{WolfKoonin:2007}. This may be due to their small sizes though even here possible solutions have been proposed (\citealt{Tamura:2015}).

It is important to note that we do not propose that the RNA world did not or could not exist, nor does this work necessarily suggest that a self-replicating RNA polymerase did not exist (although our results suggest it to be unlikely), but rather that such a molecule did not directly lead to current living systems. 
Indeed the crucial role of RNA (more correctly, XNA) in our model is highlighted by the importance of $K_R^+$, the rate of polymerisation of polynucleotide chains. We also do not dismiss any roles for ribozymes - for example it could well be that ribozymes were responsible for aminoacylation reactions (although this would inevitably raise the question of how such ribozymes were themselves replicated). Similarly, peptides alone could also have carried out supporting roles. At its core however, we suggest that the ancestral replicator was nucleopeptidic with information storage function carried out by the XNA and polymerase function carried out by the peptide.\\


 {\bf Acknowledgements}
 
 We thank Jeremy Tame, Andy Bates and Arnout Voet for critical reading of the manuscript and Arnout Voet  and Jan Zaucha for many constructive and critical discussions.
This work was funded by RIKEN Initiative Research Funding to J.G.H. And funding from the Malopolska Centre of Biotechnology, awarded to J.G.H
\newpage
\begingroup\raggedright\endgroup

\vskip 1cm
{\bf Supplementary Information}

In this section we use the SI units metres (${\rm m}$), seconds (${\rm s}$), kilogrammes (${\rm kg}$) and moles (${\rm mol}$). The mole/litre (molar ${\rm M}$) is thus expressed as ${\rm M}=10^{3} {\rm mol}/{\rm m}^{3}$ in SI units.\\

{\em Peptide hydrolysis rate}

The hydrolysis rate of a peptide bond in neutral water has been measured as 
$3 \times 10^{-9}\,{\rm s}^{-1}$, i.e. about 7 years (\citealt{KahneStill:1988}).

Hydrolysis rate of an internal peptide bond was estimated to be 
$3.6 \times 10^{-11}\, {\rm s}^{-1}$  at $25\Deg$ C, 
$1.13 \times 10^{-9}\,{\rm s}^{-1}$ at $37\Deg$ C,  
$1 \times 10^{-7}\,{\rm s}^{-1}$  at $95\Deg$ C , 
$1 \times 10^{-8}\,{\rm  s}^{-1}$  at $60\Deg$ C and  
$5.1 \times 10^{-6}\,{\rm  s}^{-1}$  at $150\Deg$ C and to be relatively insensitive to 
pH in the range $4.2 - 7.8$ (\citealt{RadzickaWolfenden:1996,Wolfenden:2014}).

So we can use values of $K_P^-$ in the range 
\begin{eqnarray*}
K_P^-:\,\, 4\times 10^{-11}\,{\rm s}^{-1} - 5.1 \times 10^{-6}\,{\rm  s}^{-1} .
\end{eqnarray*}

{\em RNA hydrolysis rate} 

DNA hydrolysis rate is pH
independent around neutral pH and in that case cleavage of
phosphodiester linkages have a half life of 140000 years (i.e. $1.6
\times 10^{-13}\,{\rm s}^{-1}$ assuming 1st order kinetics at $25\Deg$C) and 22 years 
at $100\Deg$C. RNA is much less stable,  its half life being 4 years at 
$25\Deg$ C and 
9 days at $100\Deg$ C (\citealt{WolfendenSnider:2001}). RNA is more stable
at pH 4-6 compared to higher pH (\citealt{BernhartTate:2012}) but we assume this to be a
relatively moderate effect for our suggested pH 5 and so retain
these numbers i.e. 4 years. $\approx 8 \times 10^{-9}\,{\rm s}^{-1}$. 
We actually suggest that
the system originally was XNA, possibly a mix of RNA-DNA and related
molecules, meaning that  stability was likely higher. But for simplicity it
seems reasonable to keep this estimate of 
\begin{eqnarray*}
K_R^- \approx  8 \times 10^{-9}\,{\rm  s}^{-1}.
\end{eqnarray*}

{\em Polypeptide and Polynucleotide Spontaneous Polymerisation}

Many of the parameters in our model are unknown, but we can try to estimate
them using simple kinetic theory. This will give some estimates or upper bound
on the values we should use.

In a perfect gas, the number of collisions per unit volume between two molecules 
$A$ and $B$ is given by
\begin{eqnarray*}
Z_{A,B} = \rho_A \rho_B (r_A+r_B)^2 \sqrt{\frac{8 \pi {\rm k_B T}}{\mu_{AB}}} 
\label{eq_collision}
\end{eqnarray*}
where $\rho_A$ and $\rho_B$ are the concentration of each reactant,
$r_A$ and $r_B$ the radius of the molecules, ${\rm k_B}$ the Boltzmann constant, 
${\rm T}$ the temperature in Kelvin and $\mu_{AB}={\rm m}_A {\rm m}_B/({\rm m}_A+{\rm m}_B)$ their reduced 
mass. If ${\rm m}_A >> {\rm m}_B$ then $\mu_{AB}\approx {\rm m}_B$.

If the reaction is $A+B \rightarrow C$, the equation we have to solve is
\begin{eqnarray*}
\frac{d \rho_C}{dt} = \rho_A \rho_B (r_A+r_B)^2 \sqrt{\frac{8 \pi {\rm k_B T}}{\mu_{AB}}} 
K_{A,B}
\end{eqnarray*}
where $K_{A,B} < 1$ includes the activation factor.
If $\rho_C$ and $\rho_B$ can be expressed in any units, $\rho_A$ must be 
expressed as the number of molecules per unit volume. If we express the 
densities in ${\rm mol}\,{\rm m}^{-3}$, we thus have
\begin{eqnarray*}
\frac{d \rho_C}{dt} = \rho_A\, \rho_B\, z_{A,B}\, K_{A,B}
\end{eqnarray*}
where 
\begin{eqnarray*}
z_{A,B} = N_a (r_A+r_B)^2 \sqrt{\frac{8 \pi k_B T}{\mu_{AB}}} 
\label{eq_collision2}
\end{eqnarray*}
is the quantity one needs to estimate.

For a nucleotide and amino acids we have:
 $r_{{\rm nuc}}\approx 5.5{\rm \AA}$ (\citealt{Hyeon2006}),
 $m_{{\rm nuc}}\approx 500{\rm g/mol}=500{\rm g} /6\times10^{23}\approx 8.3\times10^{-25}{\rm kg}$,
$r_{{\rm am}}\approx 5{\rm \AA}$ and
$m_{{\rm am}}\approx 100{\rm g/mol}=100g /6,10^{23}\approx 1.7\times10^{-25}{\rm kg}$
We also have ${\rm k_B}=1.38\times10^{-23}{\rm JK}^{-1}$, ${\rm T}=300{\rm K}$ so ${\rm k_BT}\approx 4\times10^{-21}{\rm J}$
and Avogadro's number ${\rm N}_a\approx6\,\times 10^{23}$.

So the collision rate of two nucleotides can be estimated as 
\begin{eqnarray*}
z_{n,n} \approx {\rm N}_a (2r_{{\rm nuc}})^2 \sqrt{\frac{8 \pi {\rm k_B T}}{{\rm m}_{{\rm nuc}}/2}} 
\approx 6.6\times10^{8}\,{\rm s}^{-1}\,\,{\rm mol}^{-1}\,{\rm m}^3.
\end{eqnarray*}

Moreover, the uncatalysed phosphodiester bond formation in solution is thought 
to have an activation energy $E_a$ of $21.1\, {\rm kcal/mol}\approx \,35 {\rm k_BT}$\,(\citealt{Florianetal:2003}). So
\begin{eqnarray*}
K_R^+= 6.6\times 10^8\, {\rm mol}^{-1}\,{\rm m}^3 \times {\rm e}^{-35}\approx 4.2\times 10^{-7}\,
{\rm mol}^{-1}\,{\rm m}^3.
\end{eqnarray*}


The collision rate of 2 polypeptides can be estimated as
\begin{eqnarray*}
z_{a,a} \approx N_a(2r_{{\rm am}})^2 \sqrt{\frac{8 \pi k_B T}{m_{{\rm am}}/2}} 
\approx 3.5\times 10^{8}\,{\rm   s}^{-1}\,{\rm mol}^{-1}\,{\rm m}^3.
\end{eqnarray*}
For the activation energy of polypeptide chains we take  
$E_{a,R}=40\,{\rm kcal/mol}=67{\rm k_BT}$ so
\begin{eqnarray*}
K_P^+= 3.5\times 10^8\,{\rm mol}^{-1}\, {\rm m}^3 \times e^{-67}\approx 2.8\times 10^{-21}\,{\rm mol}^{-1}\,{\rm m}^3.
\end{eqnarray*}

Similarly, we can estimate the collision rate between a polynucleotide of
length $L$ and a polymerase of length $\ell$ to be
\begin{eqnarray*}
z_{\ell,L} \approx {\rm N}_a (\sqrt{L}r_{{\rm nuc}}+\sqrt{l}r_{{\rm am}})^2 
\sqrt{\frac{8 \pi {\rm k_B T}}{\mu_{L,\ell}}},
\end{eqnarray*}
where $\mu_{L,\ell}=L\ell m_{\rm nu} m_{\rm am}/(L m_{\rm nu}+\ell m_{\rm am})$.
For $L=\ell=10$ we have 
$z_{\ell,L}\approx 1.5\times 10^9 {\rm s}^{-1}\,{\rm mol}^{-1}\,{\rm m}^3$.
We then have
\begin{eqnarray*}
Z_{\ell,L} \approx z_{L,l} K_{L,l}
\end{eqnarray*}
where $K_{L,l}$ is an activation factor which we have conservatively 
estimated to be $1/1500$. Moreover as all the polymerases and catalysed 
polypeptide chains do not vary much in length we can assume $Z_{\ell,L}$ to 
be independent of $\ell$ and $L$ and so
$Z=Z_{\ell,L}\approx 10^{-6}{\rm s}^{-1}\,{\rm mol}^{-1}\,{\rm m}^3$.

 \vskip 1cm
{\em RNA polymerisation rate (with concentrations)}

Experiments have been carried out with the concentration of
primordial polymerase 
(in this case an RNA molecule) present at a concentration of $2\,\mu M$, 
a template strand, present at $1\,\mu {\rm M}$ 
and a small primer (this is the RNA strand that will be extended, bound to 
the template) at $0.5\,\mu {\rm M}$ (\citealt{LawrenceBartel:2003}). 
The material to be added to the end of the 
primer, i.e. activated nucleotide triphosphates, were present at large 
excess, i.e. $100\,\mu M$. It was found that
the polynucleotide elongates by one units at a rate varying between 
$0.02\,{\rm s}^{-1}$ and $4\times 10^{-5}\, {\rm s}^{-1}$. So we have 
\begin{eqnarray*}
2\times 10^{-2}\,{\rm s}^{-1} \le k_{{\rm step}}\le 4\times 10^{-5} {\rm s}^{-1}.
\end{eqnarray*}

$h_{R}$ can be estimated from the collision rate of 2 nucleotides which we 
evaluated above to be of the order $6.6\times10^{8}\,{\rm s}^{-1}\,{\rm mol}^{-1}\,{\rm m}^3$, but we must add to that a factor taking into account the correct 
orientation of the nucleotide and so we will take as an estimate
\begin{eqnarray*}
h_{R}=10^{6}\,{\rm s}^{-1}\,{\rm mol}^{-1}\,{\rm m}^3.
\end{eqnarray*}
The actual value 
does not matter much as the limiting factor in the duplication of the 
polynucleotide is the rate $k_{{\rm step}}$.

{\em Peptide polymerisation rate}

It is estimated (\citealt{Wohlgemuthetal:2006})  that 
the rate of  peptide bond formation is approximately $0.001\,{\rm s}^{-1}$ 
under the following conditions:
at 50 mM Mg$^{2+}$, with the concentration of the primitive ribosome  (50S subunits) at $0.6\, \mu M$, 
the concentration of tRNA carrying  an amino acid (fMet-tRNA) at $6.6\, {\rm \mu M}$   and the concentration 
of the peptide acceptor (puromycin) at $10\, {\rm mM}$  (to ensure complete saturation of the ribosome).
However, it could be as small as  $10^{-8}\,{\rm s}^{-1}$ (\citealt{Sievers:2004}).

This suggests we take $k^+_{P,L}\,P_1$ in the range  $10^{-7}\,{\rm s}^{-1} - 10^{-3}\,{\rm s}^{-1} $
with $P_1\approx 10^{-5}M$, that is $k^+_{P,L}$ in the range $10^{-5}\,{\rm s}^{-1}\,{\rm mol}^{-1}\,{\rm m}^3$ - $10^{-1}\,{\rm s}^{-1}\,{\rm mol}^{-1}\,{\rm m}^3$ and
so $k^+_{P,1} \approx L k^+_{P,L}$ in the range $10^{-4}\,{\rm s}^{-1}\,{\rm mol}^{-1}\,{\rm m}^3$ - $1\,{\rm s}^{-1}\,{\rm mol}^{-1}\,{\rm m}^3$.

We have used mostly $k^+_{P,1}=0.1 \,{\rm s}^{-1}\,{\rm mol}^{-1}\,{\rm m}^3$, 
but we also considered smaller values.

\end{document}